# Compliance with restrictive measures during the COVID-19 pandemic in Europe: Does political partisanship influence behavioural responses?

Authors: Stefano Maria Iacus[1], Marco Scipioni[1,2], Spyridon Spyratos[1], Guido Tintori[1]

([1]European Commission;

[2] Honorary Research Fellowship, Department of Politics, Birkbeck, University of London)

Corresponding author: Marco Scipioni (*marco.scipioni@ec.europa.eu*)

## Abstract (150)

The success of public health policies aimed at curtailing the COVID-19 pandemic have relied on large-scale and protracted compliance by the public. A series of studies have recently argued that previous voting patterns are important predictors of such compliance. Our research further investigates such connection by tracking the relationships between parties' vote shares and mobility in six European countries over an extended period of time.

We observe that while vote shares are occasionally related to variations in mobility within each country, there is no systematic pattern of decrease or increase in mobility across all six selected countries depending on party family or government membership. Over time, the relationships between mobility and vote shares tend to grow stronger in some but not all countries, again suggesting that there is no clear connection between vote shares for several party families and compliance with social distancing measures.

# DISCLAIMER

The scientific output expressed does not imply a policy position of the European Commission. Neither the European Commission nor any person acting on behalf of the Commission is responsible for the use that might be made of this publication. For information on the methodology and quality underlying the data used in this publication for which the source is neither Eurostat nor other Commission services, users should contact the referenced source. The designations employed and the presentation of material on the maps do not imply the expression of any opinion whatsoever on the part of the European Union concerning the legal status of any country, territory, city or area or of its authorities, or concerning the delimitation of its frontiers or boundaries.

## *1* Introduction

Effective responses to the COVID-19 pandemic rely on large scale compliance with social measures imposed by governments. Our research seeks to clarify the extent to which political preferences are also related to behaviours under extreme circumstances, and thus play a role in explaining uneven levels of compliance with public health and safety measures. Better understanding whether and to what extent partisanship is related to behaviour during the pandemic remains relevant at the time of writing, as the benefits for European societies of the mass vaccination campaigns are predicated upon widespread uptake of vaccine by the population, despite pre-existing political differences.

In the US, several studies found that voting patterns at county level correlated with, *inter alia*, mobility patterns, infections, and fatality rates, and this relationship was stronger than other economic or social covariates (Goldstein and Wiedemann 2021; Gollwitzer et al. 2020). However, due to the extremely high and peculiar level of partisanship in the US, Gollwitzer et al. (2020) wondered whether their findings would replicate in other settings. In the European context, 'partisanship' - measured by self-reported vote for the incumbent government - and 'polarisation' - measured by the variation in trust towards the government - informed survey respondents' evaluations of government policies (Altiparmakis et al. 2021). In a study of three European countries, higher shares of voting for right-wing populist parties (in Denmark and Sweden) or Remain vote in the Brexit referendum (in the UK) strongly correlated with less compliance with social distancing in the aftermath of the pandemic breakout (Ansell, Cansunar, and Elkjaer 2021).

Compared to previous research, this article seeks confirmation of these insights by expanding: the remit of countries - Denmark (DK), France (FR), Italy (IT), Spain (ES), the Netherlands (NL), and Sweden (SE) -; the type of party considered - including left- and right- wing populist as well as member of government -; and covering from the breakout of the pandemic to late Spring 2021. Our key questions are: *Is compliance to social distancing during COVID-19 a partisan issue in the EU member states?* More precisely, *can we observe any correlation between vote shares and mobility patterns during the COVID-19 pandemic? And if these political differences matter, do their relevance vary over time?*

To tackle these questions, we draw on a variety of data sources. Mobility is measured by Google Mobility Data, which is anonymised and aggregated at either NUTS2 (ES), NUTS3 (ES, FR, IT), or LAU level (DK, NL, SE).[1] We include data on the last pan-European election prior to the pandemic, which is the May 2019 European Parliament (EP) election, at the same geographical level as the mobility data. We classify such voting data based on the Chapel Hill Expert Survey (CHES) (Bakker et al. 2020) to get party family (*'radical left'* or *'radical right parties'*), parties in government (extracted from Döring and Regel (2019)), as well as Zulianello and Larsen (2021)'s classification of populist parties running in EP elections. To analyse such repeated observations - as mobility in every geographical entity in our dataset has been measured daily for more than one year - we estimate separate multilevel models for each country (Gollwitzer et al. 2020). We also strive to capture variation over time by looping simple OLS regressions over time, to investigate whether the magnitude of the coefficients vary (Allcott et al. 2020; Ansell, Cansunar, and Elkjaer 2021).

In a nutshell, our answers to the previous questions are:

---

[1] For an overview of what these different geographies are, please refer to Eurostat's definitions here *https://ec.europa.eu/eurostat/web/nuts/national-structures*.

- There is some evidence that *political differences between regions or municipalities are related to variation in mobility patterns* during the pandemic in some but not all countries we consider, and this is after we hold constant other correlates of mobility such as GDP per capita, population density, age composition at NUTS3 level, employment shares in different economic sectors, as well as other pandemic-related factors such as number of cases (at NUTS3 level), and stringency of the policy measures (nation-wide).
- However, *there is no single, overarching pattern in the relationships between voting for specific political families or government and mobility across countries*, as considerable variation is recorded between them. Put differently, while one party family may be positively related to mobility in one country, it is negatively related to mobility in another. Considering the heterogeneity of the European political landscape which is captured in our data, this is not surprising. With an eye to previous studies, geographical entities where populist parties received relatively higher shares (as well as those where parties in government won most of the votes), are not systematically less (more) likely to comply with pandemic-containment measures across European countries.
- While in some countries we notice that differences in mobility between regions related to vote shares tend to become more pronounced over time, again no single pattern emerge systematically for all countries.

These results contrast with some of the previous research, which emphasised how partisanship and populist right-wing parties' votes shares were predictive of compliance with social distancing. It is difficult to compare research results across studies due to not only differences in research design and broader methodological choices, but also because of sheer measurement issues (of which we outline just a few in this paper). To be clear, we are not saying that political differences across regions do not matter. This paper argues that, after holding constant several other covariates related to, for instance, differences in population density or the intensity of the pandemic, the relationships between right- or left-wing populist parties and mobility, or between parties in government and mobility, are not consistent across countries. However, there are differences related to different party shares within countries and, within countries, over time.

The paper is structured as follows. Section 2 recaps insights from the literature and outlines the hypotheses guiding the research. Section 3 describes the data sources. Section 4 sketches the methods we use in the paper. Section 5 presents the results of the analyses. Section 6 concludes.

## *2* How does voting behaviour relate to mobility patterns during the crisis?

What are the mechanisms that may connect aggregated political preferences to compliance with pandemic-containment measures? We believe that there may be two different dynamics at play here, one regarding *ideology and behaviour*, and the second related to the *timing* of the relationship between voting patterns and behaviour during the pandemic. First, we hypothesise that party voting expresses substantial ideological differences between individuals, and that such differences may manifest into behavioural differences during the pandemic. In other words, political preferences may be consequential for the type of behaviour during a pandemic, and while voting is a complex outcome wherein many factors play a role, ideological traits are prominent among them. Second, we borrow from the literature on how public opinions are theorised to react to crises to shed some light on the variation we can expect over time. Briefly, we argue that crises do not occur in a vacuum, and that pre-existing political features are likely to become more relevant the longer a crisis persists. Put differently, while in the immediate aftermath of a crisis such as COVID-19 we would expect political differences to be compressed, the longer the crisis drags on the more likely is that such differences would re-emerge. To quickly recap, while the first expectation suggests that ideological traits may matter in behavioural terms during a pandemic, the second argues that such differences are likely to increase the more a crisis endures.

In general terms, several studies have shown a correspondence between ideological traits and behaviour during the pandemic. As mentioned in the introduction, studies focusing on the US have thought about this question in terms of partisanship in the context of a two party system. At the aggregate level, counties with higher proportions of Republican voters displayed lower levels of social distancing (Allcott et al. 2020; Goldstein and Wiedemann 2021; Gollwitzer et al. 2020), holding constant other factors such as demographic density, or confirmed number of COVID-19 cases. These insights were confirmed in survey-based studies, where again ideological leaning was found to be related to individual beliefs about the severity of the pandemic and the importance of pandemic containment measures such as social distancing (Allcott et al. 2020). However, these studies do not straightforwardly apply also to the European context, inter alia because of the differences between political systems. On the supply side, Rovny et al. (2022) have observed that party ideology was a significant predictor of European parties' stances during the pandemic. The authors also expected governing parties to be more pro-active when it comes to adoption of restrictive measures, and have higher confidence in scientific evidence. Surveys-based research has highlighted the role of opposition parties as a key element to explain differences in support and trust towards government, as well as assessments of pandemic-containment policy measures. Esaiasson et al. (2020) noticed heterogeneous increases in trust depending on political preferences of survey respondents. More precisely, while institutional trust tended on average to increase during the first months of the crisis, those opposing the governments saw more limited increases in their institutional trust on average (on this, see also Altiparmakis et al. (2021)). It can be hypothesised that voters for parties in government are more receptive towards a discourse of collective responsibility and restrictive measure to curtail the spread of the pandemic, and act more in line with governmental and scientific recommendations of social distancing.

Past studies have shown that populist voters tend to display higher distrust towards governments, which is part and parcel with the dual perception of a corrupted political elites opposed to the people.

Having low political trust towards the government has already been linked to lower levels of appreciation for governments' policies during the pandemic in studies on the so-called rally-around-the- flag (Altiparmakis et al. 2021), and low political trust is one of the defining trait of populist attitudes (Geurkink et al. 2020). Further, Ansell, Cansunar, and Elkjaer (2021) report evidence arguing that voters of populist parties are also more prone to believe conspiracy theories and fake news, which again spread fast and widely during COVID-19 pandemic. In addition, one of the essential traits of populist parties is the anti-elite stance (Mudde 2013), which may translate into lower respect for recommendations of not only political actors, but also researchers and scientists. At the aggregate level, we would expect these individual-level mechanisms highlighted in the literature to translate into lower compliance with rules and recommendations on social distancing. And indeed, Ansell, Cansunar, and Elkjaer (2021) has shown that, in the case of DK, SE, and UK, voters for populist parties display less social distancing. Ansell, Cansunar, and Elkjaer (2021) reached those conclusions only relative to right-wing populist parties because of the specific countries the authors selected, which happened to feature only such form of populism. Thus, it is not clear whether what Ansell, Cansunar, and Elkjaer (2021) observe is a common feature connecting areas where populist parties received higher than average vote shares during the pandemic, or whether that is limited only to right-wing populist parties. At a higher level, and looking at the supply side, Rovny et al. (2022) found substantial differences in political parties' responses to the pandemic depending on left/right stances, but it is not clear whether this applies also to different forms of populism and whether it also matches public behaviour in areas where these parties received relatively higher shares of votes.

In the light of the above, our first hypotheses are:

- H1a) The higher the vote shares for cabinet parties, the higher the reduction in mobility.
- H1b) The higher the vote shares for populist parties, the lower the reduction in mobility.

Turning to our second expectation, a common framework for analysing public opinion reactions to crises in general and the COVID-19 pandemic in particular is the so-called *rally-around-the-flag* (Newman and Forcehimes 2010). Such framework has not only repeatedly appeared on the media to explain the surge in popularity in the early weeks of the pandemic (Economist 2020) but has also been analysed extensively in the academic literature (Kritzinger et al. 2021). The basic idea is that people will tend to support their leaders no matter the previous beliefs, ideological orientations, or partisanship, in the context of national crises such as wars or natural disasters. The observable implication of this framework is that pre-existing political differences should shrink as a consequence of a crisis. The empirical evidence supporting such claims is mixed though. While in a cross-national study, Bol et al. (2020) find that the rise in support spans across the left-right ideological continuum, other studies do observe uneven levels of support along political lines (Altiparmakis et al. 2021; Kritzinger et al. 2021; Oana, Pellegata, and Wang 2021). Kritzinger et al. (2021) suggest that we should think about rallies-around-the-flag as short outburst of support from a very heterogeneous coalitions of groups, including voters of opposition parties. The longer the crisis drags on, the higher the chances are that several groups will withdraw their support. Altiparmakis et al. (2021) explicitly hypothesise that, during the pandemic, policy evaluations of government measures should strongly diverge based on trust and partisanship the more the crisis drags on.

On this basis, we formulate the following hypothesis:

- H2) Differences in voting patterns should correlate with different behaviours during the pandemic only gradually and over time.

## 3 Data

From a data perspective, one of the added value of this paper is that, contrary to much of the literature emerged so far, we can differentiate between different waves and timings during the pandemic. Google Mobility Data starts on 15/02/2020, and we stop tracking it between April and May 2021 depending on the country. The studies that so far have the longest coverage stretch into the summer of 2020 (Ansell, Cansunar, and Elkjaer 2021; Kritzinger et al. 2021). Both the mobility and the political data is at the same geographical level - i.e. NUTS3 or LAU. As with all geographically aggregated data, our observations are limited by the fact that we do not observe individual level data, and therefore our conclusions should not be necessarily regarded as valid also for that level.

### 3.1 Mobility data

In this paper, mobility during the pandemic is taken as a proxy of compliance. This is related to the fact that many countries have adopted measures strictly limiting population movement during the pandemic. In a somewhat looser fashion, the overwhelming scientific consensus and recommendations by governments was to limit unnecessary movement as much as possible and, as a consequence, the possibility of contact and transmission. To be clear, mobility here is a proxy, not a direct measurement of compliance. Ideally, compliance with government policies should be measured at individual level - i.e. whether individual $i$ at time $t$ in place $p$ was abiding or not to policies aimed at stopping the pandemic. In absence of such data, we use mobility at regional or municipal level to understand whether, in aggregate, people were moving less, thus providing an indication on whether policies were respected in those regions.

To our understanding, Google Mobility Data takes as baseline roughly the median for each weekday during January 2020, and then normalises subsequent mobility based on that reference value. What kind of mobility is actually measured by Google is clarified as follows: 'We calculate these insights based on data from users who have opted into Location History for their Google Account, so the data represents a sample of our users. As with all samples, this may or may not represent the exact behavior of a wider population'.[2] To get rid of some of the noise due to daily fluctuations, we aggregate the daily data into weeks by taking both the mean of the mobility indicators for each geographical entity and week (Mon-Sun). It is important to appreciate that *results based on such normalised indicators should always be interpreted as relative increases (or decreases)*, and *not in absolute terms*. This is not a trivial point as we have no way to know - absent a time series also covering prior years - how representative the baseline period is of the mobility in a given locality, as also Google[3] explicitly acknowledges regarding its Mobility Data.

---

[2] *https://www.google.com/covid19/mobility/data_documentation.html?hl=en*.

[3] Google's data overview states: 'How did we pick perfectly normal baseline days? We probably haven't—a short period of the year can't represent normal for every region on our planet. We picked a recent period, before widespread disruption as communities responded to COVID-19.

Figure 3.1 shows the descriptive trends for the average weekly retail mobility (top row) and an aggregate formed by mobility related to grocery, retail, transit, and workplace, as recorded by Google. Mobility actually increased approximately during summer 2020 in all Member States, to then decrease again during late Autumn/early Winter. That being said, the intensity of the mobility contractions, as well as specific timing of changes, vary between Member States and typologies of movement. In our analysis we decided to drop the *residential* and *parks* mobility, as it showed patterns that were either just opposite to all others typologies (with the exception of Sweden), or difficult to reconcile conceptually with the other mobility types.

We expect to see greater reductions in mobility in *retail* as compared to *grocery* and others. The former can be regarded as something including a greater share of non-essential movements, whereas the latter are much more difficult to decline as people still have to shop for essential goods also during a pandemic. We produce one aggregate form of mobility - henceforth, *aggregate mobility* -, which averages all types of mobility except for residential and parks. In subsequent plots and analyses, we show the results for the aggregate indicator, and leave all results relative to retail in the Appendix. Analyses related to other forms of mobility are in the GitHub repository. While we would expect more differences based on the retail category, the

---

Even so, for some regions, the baseline falls during a time when COVID-19 was established. To interpret the data for your region, follow the local checklist', available at

[https://support.google.com/covid19-mobility/answer/9824897?hl=en&ref_topic=9822927](https://support.google.com/covid19-mobility/answer/9824897?hl=en&ref_topic=9822927).

aggregate indicator we produce has fewer data gaps and missing[4], as there are days or weeks where we simply lack data for specific indicators in some geographical entities. Further, it is not entirely clear how the mobility typologies are defined, what counts as mobility for one category but not others (e.g. what counts towards 'transit' and not 'work'?), the degree to which they overlap, or how specific pandemic containment measures in different countries may have affected individual mobility types (e.g. closure of pubs, restaurants, or limitations on opening hours; on this, see also Lipsitz and Pop-Eleches (2020)).

---

[4] Indeed, about a third of the unprocessed Google Mobility Data is NA, namely 1221777 out of 4070406 observations.

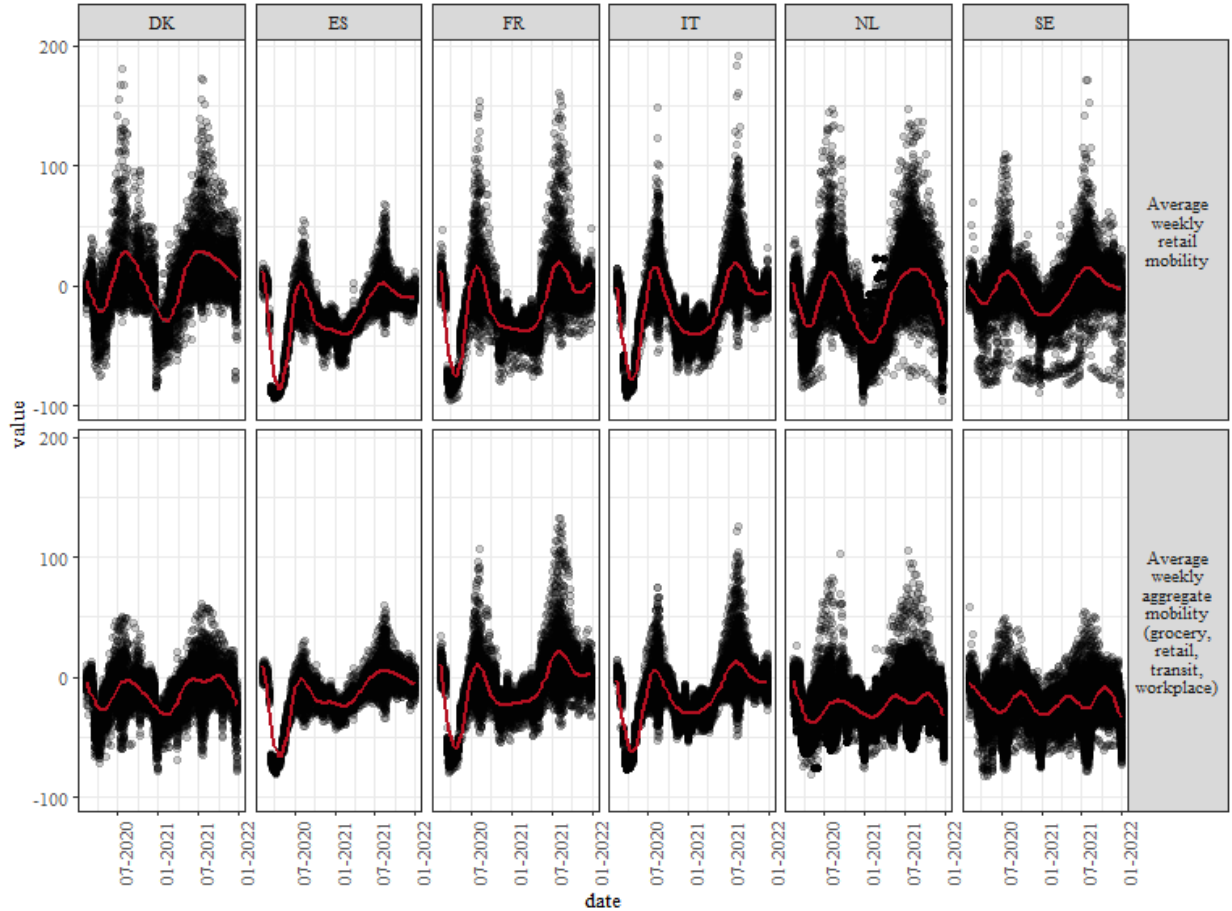

*Figure 3.1: Retail and Aggregate Mobility*

## 3.2    Data on COVID-19 confirmed cases and policy measures

Considering that our outcome variable is either at NUTS3 or LAU level, the best case would be to have all other variables at that level. Unfortunately, that is not always possible, as important information such as confirmed COVID-19 cases or policy measures may not be available at different geographical level or only with a scattered coverage.

*3.2.1* **Regional data on confirmed COVID-19 cases**

We use the daily cumulative confirmed cases at NUTS3[5] level from the dataset assembled by Guidotti and Ardia (2020). Using the cumulative daily cases, we estimate the weekly number of new confirmed cases by 100K inhabitants for each region. The time series of confirmed cases start in different moments for the countries we consider in this paper. For Spain, the time series start on 1st January 2020, in Sweden on 4 February 2020, in Italy on 24 February 2020, in France on 13 May 2020, in Netherlands on 28 February 2020 and finally in Denmark on the 1st of March 2020. Thus, in France we lose the entire so-called *first wave* of the pandemic. We notice substantial heterogeneity in both intensity of the epidemiological curve and its timing across countries (Figure 3.2). In addition to the differences across countries, we also notice substantial within-country variation for the six countries we selected.

---

[5] Due to some inconsistencies between how the COVID-19 dataset were created in Guidotti and Ardia (2020) and Eurostat classification, we lose some regions in a few countries.

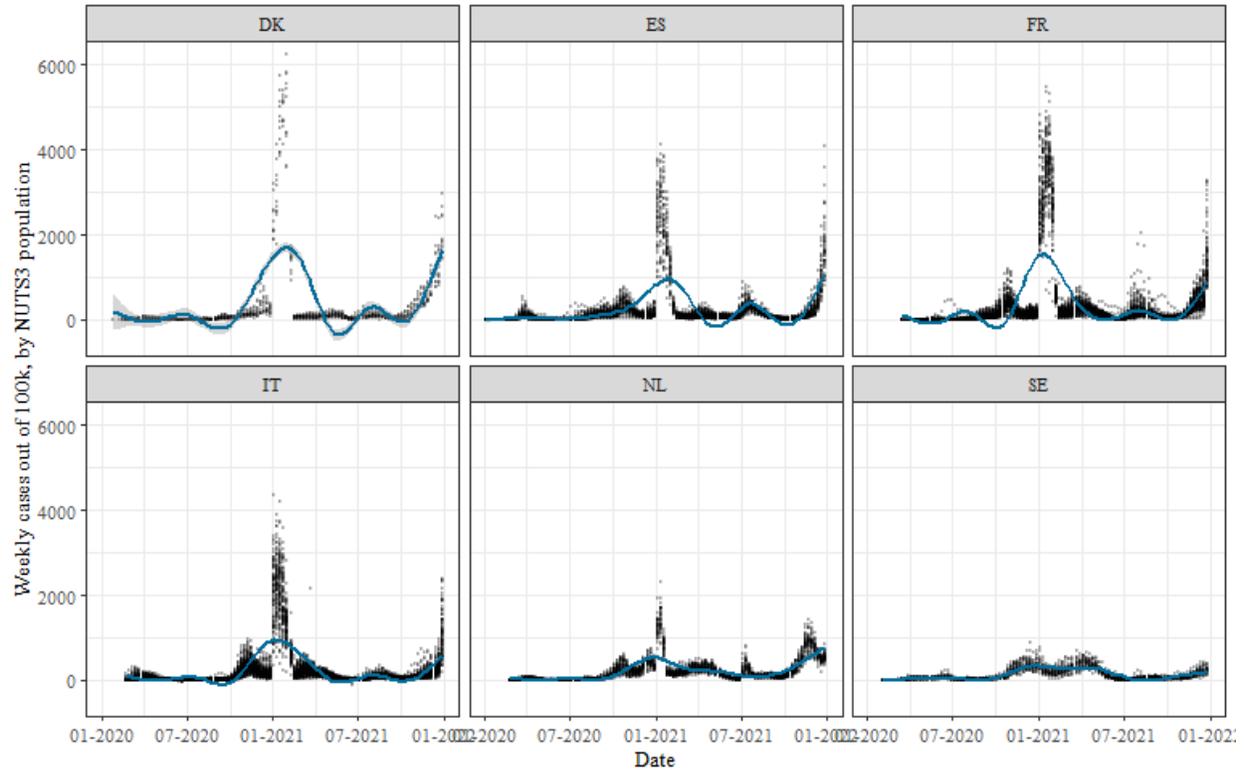

*Figure 3.2: Weekly cases over 100k population, by country and week. Dots are NUTS3. Blue line is a smoothing spline highlighting the trend line across all NUTS3 for each country*

### 3.2.2 Oxford COVID-19 data on policies

To include information on policy measures we turn to the data from the Oxford COVID-19 Government Response Tracker (OxCGRT) (Hale et al. 2021). We selected the Stringency Index - which is an aggregation of several underlying indicators - because, even if there are specific items focusing on movement restrictions (e.g. '*c stay at home requirements*', '*c7 movement restrictions*'), mobility could be affected also by other policy measures such as closing of schools, or restrictions on workplaces, or the use of public transport. Our analysis reveal that there is a strong and negative correlations between mobility and Oxford's Stringency Index. This is expected, as the higher the stringency of the measures being adopted, the less the mobility. Figure 3.3 plots the daily variation in the daily stringency index by country. There is great heterogeneity between countries in terms of both intensity of the measures being adopted, as well as their timing. That said, one common pattern is that by approximately June 2020 all countries had enacted at least a first batch of stringent measures. The exception to this pattern is Sweden, which not only implemented relatively milder measures, but also tends to display less variation over time.

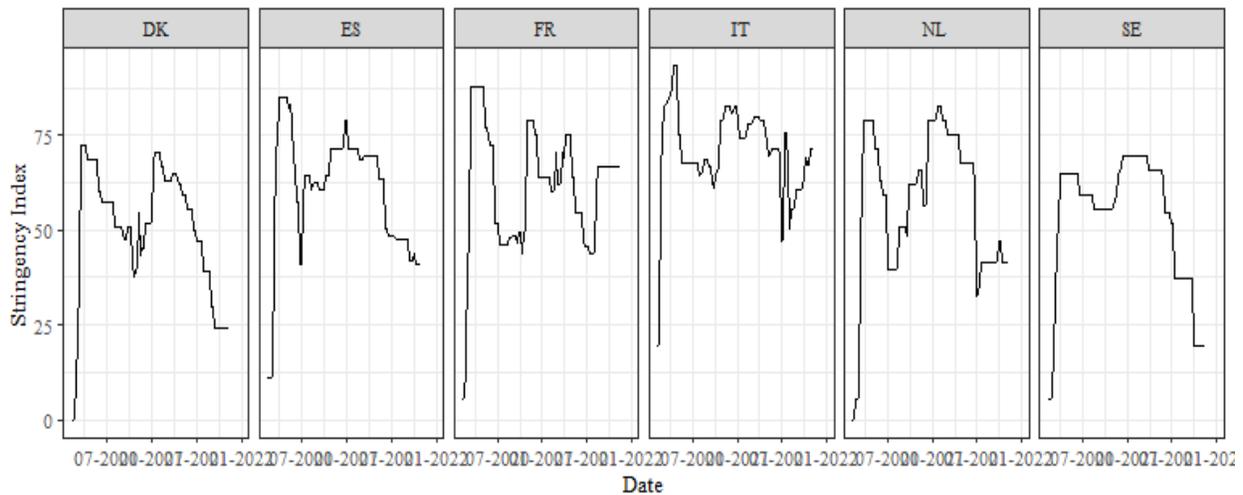

*Figure 3.3: Stringency Index, daily and by country*

## 3.3 Political data

Our main independent variable is the political parties' vote shares at NUTS3 or LAU level in the last EP election held in May 2019. That election immediately preceded the outburst of the pandemic in early 2020. The sources of the electoral data for each of the MS included in the dataset are the official electoral bodies of the country. The original electoral data was collected at the lowest electoral geography available (e.g. polling station, municipalities), then aggregated and georeferenced at LAU level using the 2019 LAU grid by Eurostat. We then aggregated up this LAU level data to NUTS3 for France, Italy, and Spain.

We connect political parties listed in the national data to their CHES classification (Bakker et al. 2020). In the CHES data populism is not explicitly listed as a political dimensions, but there are other dimensions which are traditionally held to be amongst its defining traits (Mudde 2013).

More precisely, we check three political dimensions from CHES - namely, *anti-elitism*, *anti-elitism salience*, and *salience of political corruption* -, and verify the party selection according to the *party family* listed always in CHES, as well as the separate classification of populist parties in EP elections provided by Zulianello and Larsen (2021) (Figure 3.5).[6] For the remainder of this article, we will use the CHES party family classification - for our purposes, the *far-left parties* and *far-right parties* - as checked against Zulianello and Larsen (2021). To facilitate comparisons with Ansell, Cansunar, and Elkjaer (2021)'s results, we adopt PRL (populist radical left) and PRR (populist radical right) as labels in the analysis. In addition, we also include dummies capturing whether a political party was a government member or not, extracting such information from Döring and Regel (2019). We believe that it is important to compare different party families, as well as simpler government/opposition dynamics, to capture the extent to which the patterns we witness are peculiar to one party family or something cutting across the political spectrum.

This selection results in the following parties.[7]

- In Denmark, there is only one party classified by CHES as *radical left*, namely *Enhedslisten—De Rød-Grønne Unity* (List-Red/Green Alliance, EL; 4.18% of the vote nationally in the 2019 EP election), and another party that is classified as *radical right*, that is *Dansk Folkeparti* (Danish People's Party, DF; 10.8%). While the former is not included as a populist 'leftwing' party in Zulianello and Larsen (2021), the latter is listed as both 'rightwing' and 'radicalright'. The cabinet is composed exclusively by *Socialdemokraterne* (Social Democrats, SD, 21.5%).

---

[6] See also the *Case selection* sub-section for more details.

[7] See table in the appendix for reference.

- There are two parties classified as *radical left* in France by CHES, DLF (*Debout la France*, 3.51%) and RN (*Rassemblement national*, 23.34%). These parties are also classified as populist *right wing* and *radical right wing* respectively by Zulianello and Larsen (2021). On the opposite side of the spectrum, *Lutte Ouvrière* (LO, 0.78%) and *La France Insoumise* (LI, 6.31%) are classified as *radical left* in CHES, but only the latter is also classified as populist *left-wing* party in Zulianello and Larsen (2021). The government was composed of *La République En Marche* (LREM, 22.42%) and *Rassemblement pour la République / Union pour un Mouvement Populaire / Les Républicains* (RPR/UMP/LR, 8.48%)

- There are two parties classified as populist in Zulianello and Larsen (2021) in Spain: the alliance between *Podemos* and *Izquierda Unida* (*Unidos Podemos*, 10.1% of the vote in the 2019 EP election), which is classified as *left-wing*; and *Vox* (6.2%), which is classified as both *right-wing* and *radical-right*. The former is classified as *radical left* in CHES, while the latter is classified as *radical right*. The government was composed of *Podemos* and *Partido Socialista Obrero Español* (PSOE, 32.86%).

- In Italy, the *League* (LN, 34.26%) is classified by both CHES and Zulianello and Larsen (2021) as a far-right and right-wing populist party. Instead, *Brothers of Italy* (FdI, 6.44%) is not defined as a far-right party by CHES but just a right wing party, whereas it is classified as both 'right-wing' and 'radicalright' by Zulianello and Larsen (2021). There is no *far left* party which got a seizable amount of votes in the 2019 EP election, and Zulianello and Larsen (2021) classification does not feature any populist party on the left.

- *Movimento 5 Stelle* (M5S, 17.06%) is classified as belonging to 'no family' by CHES, and a 'valence' populist party by Zulianello and Larsen (2021). M5S was also a government member, together with *Partito Democratico* (PD, 22.74%) and *Sinistra e Libertà* (SEL, 1.75%).

- In the *Netherlands*, there are two *radical right* parties in CHES which are also included in Zulianello and Larsen (2021) as both 'rightwing' and 'radicalright', namely *Forum voor Democratie* (Forum for Democracy, FvD; 11%) and *Partij voor de Vrijheid* (Party for Freedom, PVV; 3.5%). While no party is classified as 'leftwing' in Zulianello and Larsen (2021), CHES classifies as *radical left Socialistische Partij* (Socialist Party, SP; 3.37%). Two parties composed the cabinet, namely *Volkspartij voor Vrijheid en Democratie* (People's Party for Freedom and Democracy, VVD; 14.64%) and *Christen-Democratisch Appél* (Christian Democratic Appeal, CDA; 12.18%).

- In Sweden, CHES classifies *Vänsterpartiet* (Left Party) as *radical left*, but that party is not classified as populist *left-wing* in Zulianello and Larsen (2021). CHES categorises the *Sverigedemokraterna* (Sweden Democrats) as *radical right* party, and Zulianello and Larsen (2021) classifies that party as populist *right-wing* and *radical right*. *Arbetarpartiet-Socialdemokraterna/Sveriges Socialdemokratiska Arbetareparti* (Worker's Party-Social Democrats and Swedish Social Democratic Party, SAP; 23.5%) and *Miljöepartiet de Gröna* (Environment Party—The Greens, MP; 11.5%) were part of the government.

## 3.4 Other covariates of interest

Previous studies have shown the importance of including in the analysis information on other factors that may be related to overall levels of mobility, such as demographic or economic aspects of a given geographical entity (Goldstein and Wiedemann 2021; Ansell, Cansunar, and Elkjaer 2021; Gollwitzer et al. 2020). We follow this lead and include: *GDP per capita*; *share of employment* in 'agriculture, forestry and fishing', 'Industry', 'Manufacturing', 'Construction', 'Wholesale and retail trade; transport; accommodation and food service activities'[8]; *old age dependency ratio*; *demographic density*. These covariates are all at NUTS3 level, and extracted from Eurostat databases.[9].

We would ideally opt to include more information on a number of other regional characteristics, such as education or relative wealth of a region. Unfortunately, some of these variables are either not present in a harmonised manner for all countries, or are only available at higher geographical resolution (e.g. education is only available at NUTS2 level).

## *3.5* Case selection

We select Denmark, France, Italy, the Netherlands, Spain, and Sweden. First, these are large enough countries with a minimum number of NUTS3 or LAUs allowing us to explore regional differences. Denmark has 99 LAUs, France has 101 NUTS3 (*Départements* and *DOM* in national administrative units), Spain has 59 NUTS3 (*Provincias, Islas, Ceuta, Melilla*), Italy has 107 NUTS3 (*Provincie*), the Netherlands has 355 LAUs, Sweden has 290 LAUs (*Län*).[10] Second, as

---

[8] Unfortunately, such information is missing for Sweden.

[9] More details are in the GitHub repository

[10] For an overview, see *https://ec.europa.eu/eurostat/web/nuts/national-structures*.
Unfortunately, due to boundary changes, missing values in the original Google Mobility Data, we do not have a complete coverage of all countries. In Italy, we lose almost all NUTS3 in Sardinia

Figure 3.4 shows, there is substantial heterogeneity within and between countries along all the demographic or economic aspects listed in the section above.

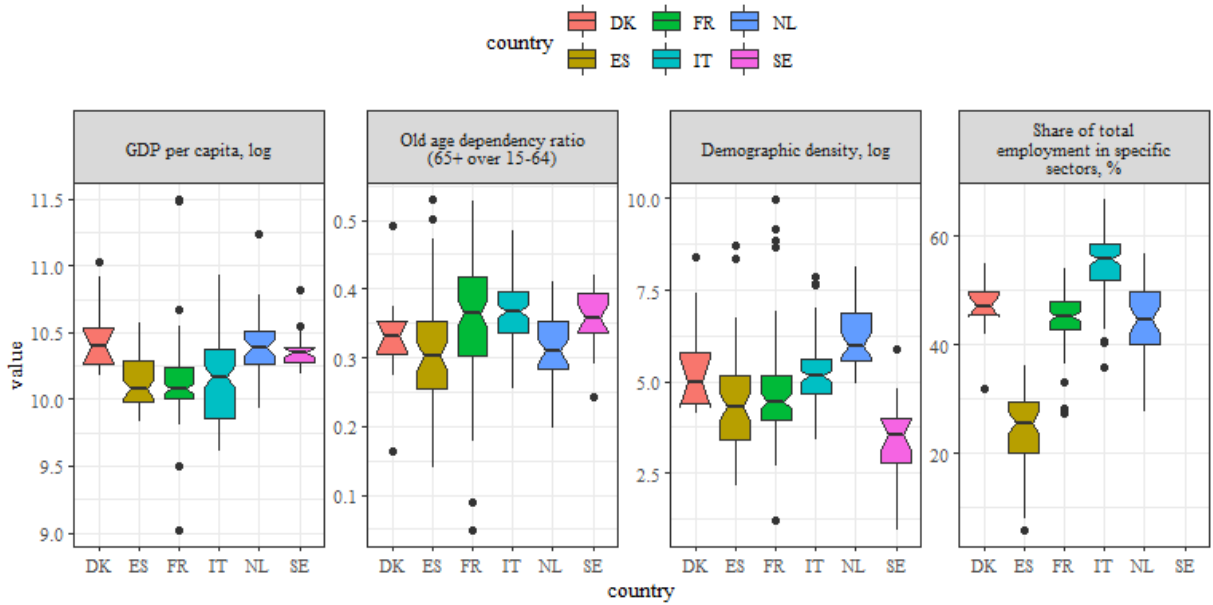

*Figure 3.4: Distribution of NUTS3 by covariates and Member States*

Figure 3.5 captures only part of this variation when it comes to political parties' families and their scores on three dimensions likely to capture some of the defining features of contemporary populism in Europe, namely their *anti-elitism*, the *salience of such anti-elitism*, and the *salience of political corruption*. We can notice that two far-right parties such as *RN* and *LN* score rather high on at least 2 key dimensions of antielitism, but less so on political corruption. However, another far-right party such as *VOX* scores very low on the *People v Elite* dimension. Figure 3.5 suggests that while there are commonalities among populist parties in the six Member States, there are also substantial differences (and the left-right dimension is just one of them).

---

due to boundary changes, in DK, NL, SE, we lose 5, 5, and 19 respectively due to missing values in the original Google Data.

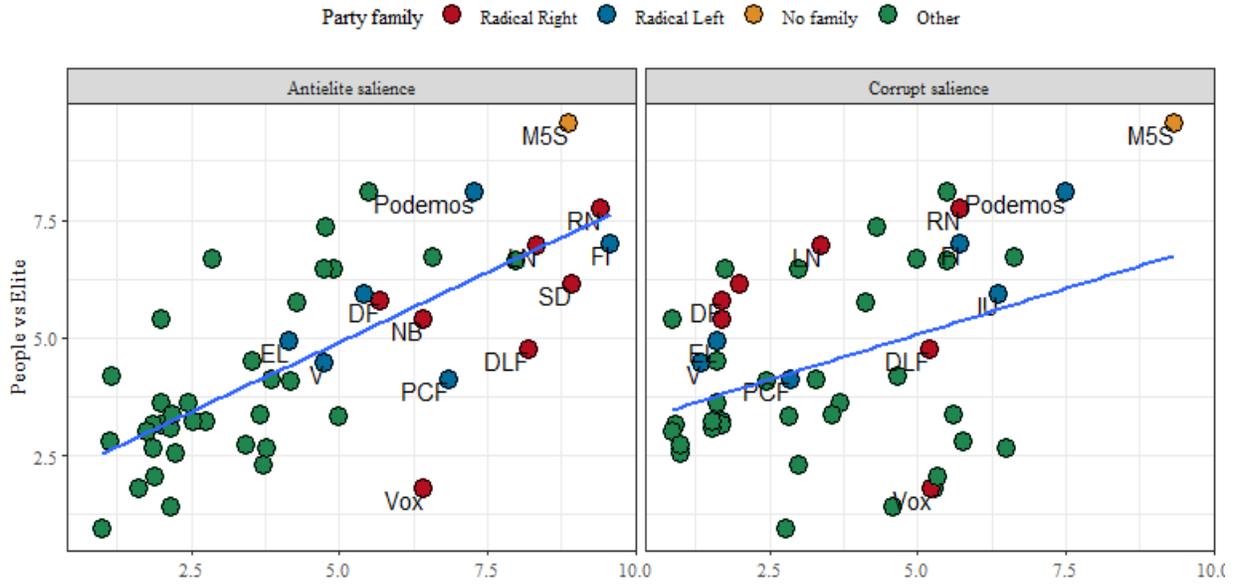

*Figure 3.5: Political parties along several CHES dimensions*

Third, as Figures 3.3 and 3.2 show, these countries display some variation in both the stringency of the measures being adopted as well as the intensity of the epidemiological situation. Fourth, a key purpose of the paper is to explore whether political differences at the sub-national level are related to compliance with pandemic-containment measures, here exemplified by reduction in mobility. In the only other European study we are aware of, Ansell, Cansunar, and Elkjaer (2021) wanted to focus on whether the anti-elitism of populist parties translated into scepticism towards governments' policy making during the pandemic. Due to the fact that the authors selected Denmark, Sweden, and United Kingdom, they ended up with only right-wing populist parties. Thanks to our country selection, in our investigation we can expand this by focusing on not only right-wing populism, but also left-wing.

# 4 Method

We agree with Berry et al. (2021)'s statement that due to the complexity of the crisis, the difficulty in obtaining reliable measurements in both outcomes (how representative is the sample of people activating geolocation on their mobile phones?) and several predictors (what is a good measurement of the intensity of the crisis? how do we measure policy change in the context of a pandemic?), perhaps the best strategy is not to rely on just one modelling specifications, but make several attempts with reasonable modelling choices.

To capture the repeated measurements in the data, we estimate a series of multilevel models (Roback and Legler 2021)[11]. To better explore how political ideology is related to mobility, we interact the former with dummies for periods - under the assumption that the relationships between ideological leaning of a geographical entity and mobility may vary over time during the pandemic. To make sense and present the hierarchical models, we plot the predicted values of models with both the average and normalised indicators as outcome variables, according to best practices in the field (King, Tomz, and Wittenberg 2000).

In another effort to capture the time dimension in our analysis, we loop OLS regressions over all weeks in our data for each country. This gives an idea of whether a coefficients changes in magnitude over time, with the ultimate goal of verifying whether political differences become more important over time. This approach is already displayed in Allcott et al. (2020) and Ansell, Cansunar, and Elkjaer (2021), and here we replicate it.

We should make clear that our analysis does not aim at and is not suited for identifying causal mechanisms - in other words, it is a correlational analysis.

---

[11] With the *lme4* R package (Bates et al. 2015). Regression equations and tables are in the Appendix.

# 5 Results

In the following sections, for each Member states, we proceed in a stepwise fashion. We first present the plots of predicted probabilities capturing the main interactions of interest stemming from the multilevel models.[12] Second, we display the plot of the main coefficients of interest from the OLS looped over each week in our data.

## 5.1 Changes over time

### 5.1.1 Predictions from interactions featuring pandemic periods with parties' shares of votes

We first create a series of dummies marking substantial policy changes occurred in the country[13]. Then, we run multilevel models which features repeated observations over time and where political variables are interacted with the aforementioned dummies capturing policy changes. In order to better interpret the results, we show the predicted values (Figure 5.1) of the models based on different levels of the variables included in the interactions, holding all other covariates constant at their means. The graphs are constructed as follows. The observed party shares are on the horizontal axis, whereas the predicted mobility values are on the vertical axis. Both axes are not fixed, as party shares and changes in mobility may have very different ranges. Facets are ordered by countries (first label on top), and for each country by party family (second label starting from the top). The lines are coloured depending on the pandemic periods (based again on those dummies marking noteworthy policy changes). These are labelled '*0*' for the baseline period, '*1*' for the period roughly starting in March when the first measures of containment were put in place; '*2*' for the period roughly corresponding to the summer and when the first pandemic containment measures were partially lifted;[14] '*3*' for the period starting in Autumn were restrictive measures were re-introduced. France also has a period *4* (but not a first, due to data gaps in the collection of confirmed cases, as described above in the Section 3), Spain does too, and the Netherlands has also a fifth. To reiterate, the dates marking these periods are different for each country, as the pandemic and the policy response to it have had different courses across Europe. We are aware that all choices regarding periodisations are somewhat arbitrary. Therefore, we provide alternative ways to capture the timing of the pandemic and its policy response in the Appendix. These alternatives are more data driven, being related to local minima in the trend lines of both confirmed cases and Stringency Index.

---

[12] More details on the models are in the Appendix.

[13] See Appendix and country scripts on the GitHub repository for more details

If the hypotheses outlined in Section 2 were correct, we would expect to see in Figure 5.1:

- H1a) lines trending down as shares of votes for parties in government increase;
- H1b) lines trending up as shares of votes for populist parties increase (with the possibility of further differentiating between left- and right-parties);
- H2) slopes of lines changes depending on the period, as over time differences tend to increase. On the contrary, if lines are parallel, it means that mobility may have changed during the different pandemic periods, but it has changed in the same manner disregarding the political differences between geographical entities.

---

[14] In both Figures 5.1 and 8.15 we omit to display the prediction for this waves, as it would unnecessarily complicate the plot. In any case, this period was one of heightened mobility due to the lifting of restrictions in all countries analysed in the paper.

Amongst the countries we analyse, no systematic patterns of support for either H1a or H1b emerge (Figure 5.1). The predicted values relative to vote shares for parties in government (first column on the left) only decrease in the case of Spain (very mildly) and the Netherlands (and even there not in all periods). The same pattern emerges also when we look at retail mobility (Figure 8.15). Denmark and the Netherlands offer support towards H1b) when it comes to populist right-wing parties (Figure 5.1, right-hand column). If we look at retail mobility, we notice also a steep upward slope also in Sweden (Figure 8.15), especially during period 1, thus confirming Ansell, Cansunar, and Elkjaer (2021)'s insights. This, however, is not confirmed when aggregate mobility is the outcome (Figure 5.1, suggesting large variation by mobility type. However, in the Netherlands the slopes for periods 3 and 4 become approximately flat in the case of retail mobility. We again do not observe a pattern systematically supporting the hypothesis when it comes to populist left-wing parties.

Turning to H2), The case of PRL parties in ES during the period 3 is an exception as it clearly shows a divergent trend compared to the other lines in the same facet (Figure 5.1). Another interesting exception is period 3 in Italy. That embraces roughly the so-called *second wave* in the country, and we could see the line becoming flat no matter the political party. This reflects the fact that while the so-called *first wave* mainly concentrated in the northern regions where the *LN* historically has had its strongholds and left the south of the country relatively unaffected, during the so called *second wave* such geographical differences in the spread of the pandemic disappeared. The important point is that, if party shares mattered, we would expect lies to bend in different directions, which is not the case. The Netherlands offers perhaps the most variation both across parties and periods. There are also large changes in the intercepts, signalling substantial oscillations in overall mobility across periods. Turning finally to Sweden, while the pictures for both cabinet and PRL parties seems one of little change over time, in the case of PRR parties there is a stark contrast between periods 1 and 3, with mobility sharply decreasing the higher the PRR vote share during the so-called *wave 1* (period 1 in blue), whereas the relationship becomes flats with the so-called second wave (period 3 in purple). It is also interesting to notice that we do not observe substantial differences between periods 1 and 2, which is in line with the fact that the trend lines over these two periods in Sweden do not exhibit the same steep declines as in other countries (see Figure 3.1).

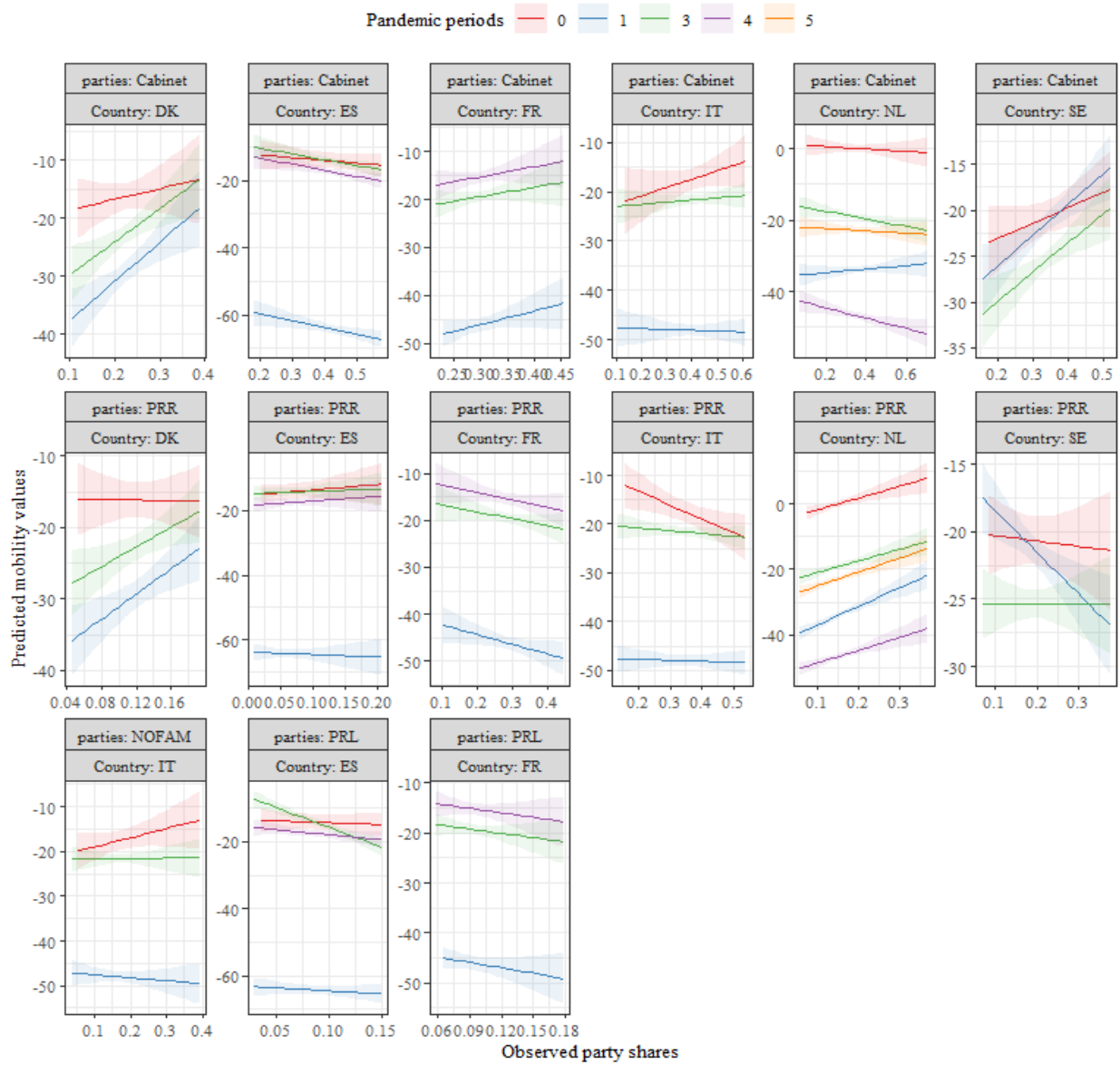

*Figure 5.1: Predicted values of aggregated mobility, at different levels of parties' share, by different waves and countries.*

### *5.1.2* **Weekly analysis**

As a second effort at capturing over time variation, we loop OLS regressions over each week to visually investigate whether coefficients change over time (similar to what Allcott et al. (2020) and Ansell, Cansunar, and Elkjaer (2021) do). The clear advantage of this alternative approach is to have a more granular appreciation of changes over time, as each week in the observational period is displayed. On the contrary, the approach based on dummies relative to noteworthy events - detailed in the previous subsection - pools all data over relative long stretches of time, therefore averaging over what may be important differences.

Similar to the subsection above, we ask ourselves that these plots should look like were our hypotheses correct.

- H1a) coefficients for parties in government should be below zero;
- H1b) coefficients for populist parties should be above 0;
- H2) coefficients should get further away from 0 (and in the direction expressed in H1a and H1b) over time.

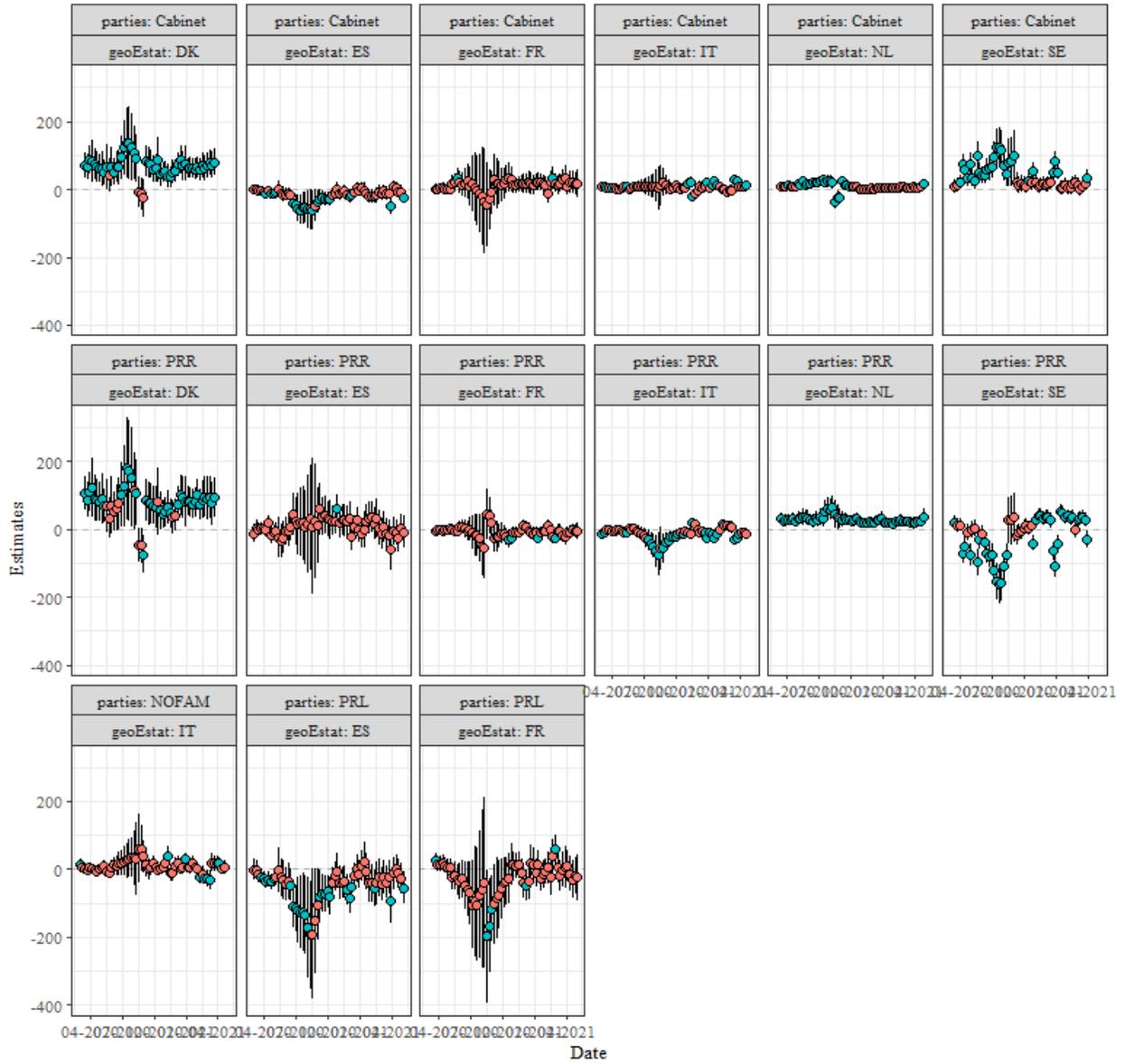

*Figure 5.2: Weekly coefficients by country and party family. Dots are coloured depending on whether p<0.05 (blue) or not (red).*

The only country where we see some support for H1a is Spain, and even there only until the end of summer 2020 (in both cases of aggregate and retail mobility, Figure 5.2 and 8.16 respectively). Turning to H1b, we can partly confirm Ansell, Cansunar, and Elkjaer (2021) findings regarding PRR in Denmark, as we observe a positive relationship between PRR vote share and aggregate mobility (Figure 5.2). However, we also qualify these remarks in two important ways. First, thanks to the comparative lenses adopted in this paper, we can also observe that parties in the cabinet show a similar pattern in the case of aggregate mobility (in the case of retail mobility most coefficients include 0 though, Figure 8.16). Second, this finding is not constant over time if we look at retail mobility, as the estimates get closer to 0. The other two countries were we see a confirmation of H1b are the Netherlands and, with substantial gap during summer - Sweden, but only if we look at the relationship between aggregate mobility and vote shares. If we switch to retail mobility (Figure 8.16), these relationships are no longer there for Sweden, and only up until the beginning of the summer in the Netherlands.

Concerning H2, the most striking case is the Netherlands. The two different types of mobility, aggregated and retail, provide a radically different picture for the Netherlands. Similarly to Denmark, PRR parties in the Netherlands show positive relationships with aggregate mobility, which is constant over time. However, if we turn to retail mobility, the coefficient pivots during summertime from markedly positive to markedly negative. Always looking at retail mobility, we observe the opposite patterns for cabinet parties, whereby the coefficients switch during summertime from negative to positive. On the other hand, the only period in which the relationships between cabinet party shares and aggregate mobility is statistically significant is exactly summertime.

Finally, in the case of Sweden, PRL and PRR coefficients seem to move in opposite directions until the summer of 2020 (respectively, positive and negative), but afterwards differences seem to shrink and both coefficients tend to be estimated around 0. This is limited to aggregate mobility though, as in the case of retail mobility there does not seem to be any difference between the shares of the two party families, as both hover around 0.

# 6 Conclusions

This article set out to understand whether partisanship can help to understand differences at aggregate level in compliance with pandemic-containment measures between EU regions. While there have already been several attempts at using partisanship to explain behavioural differences during COVID-19 in the US (Gollwitzer et al. 2020; Goldstein and Wiedemann 2021), there is only another study that we are aware of that has a comparative approach in the European context (Ansell, Cansunar, and Elkjaer 2021).

There are several noteworthy findings from this research. First, there seems to be *no systematic relationship across countries* between voting patterns - be they for parties in government, or populist parties - and mobility, let alone any differences in pattern between right- and left-wing populism. More precisely, while there is some evidence that some regions where populist parties received relatively higher shares of votes increased their mobility in some of the countries in our sample, there is no consistent pattern across all countries in our study. Similarly, vote shares for government are not systematically related to lower mobility patterns.

Another guiding expectation of this paper was that, while it is plausible that in the early weeks and months of the pandemic a sort of panic effect could have dominated the initial reaction (Schraff 2020), as the pandemic dragged on, the heterogeneous coalitions that rallied to support the government were likely to dismantle (Kritzinger et al. 2021). Again, there is mixed evidence supporting that conjecture. Roughly starting with the so-called *second wave* in Europe (Autumn/Winter 2020), differences in shares of votes for PRR or PRL parties became more relevant in mobility patterns in the Netherlands and Sweden, but this depends upon the type of mobility we select and does not travel well across them.

The paper gas several limitations. First, we analyse compliance, which is an individual attribute, with aggregate data at regional or local level. The risks for ecological fallacy are real, so the results we present should be verified also at the individual level. In addition, the level of measurement matters. We strove to always use the most detailed geographical data at our disposal, but it is plausible that our findings may be challenged by studies leveraging finer-grained geographical data. Also, mobility is a noisy proxy for compliance, as the former is a complex phenomenon determined by a multiplicity of factors. For instance, we have no local-level information on sanctioning, or even effectiveness of public administrations at local level, which we could use to address some of the concerns regarding omitted variable bias. A second limitation is that we have no comparison point in terms of behaviour prior to the pandemic. This is likely to be another major limitation as we cannot compare how mobility patterns changed before and after the pandemic, but only between regions over time since the pandemic started. More in general, mobility as recorded by Google is a black box which we have no way to verify or pick apart. In addition, sub-national political divisions may overlap with chronological differences in the geographical spread of the pandemic, to the point of making relative attribution of significance problematic (this is particularly evidence in the case of Italy).

That being said, we still believe that our study speaks to important issues in public debates. Political differences at the sub-national level are related to differences in mobility patterns, but the same is true for other factors. Indeed, it would be surprising that a life-changing event as a pandemic does not trigger behavioural differences connected to political identities. Furthermore, considering that the pandemic has morphed into a protracted crisis, affecting almost every aspect of anybody's life, it is also very likely that political differences become more relevant as time goes by. In addition, our paper underlines that the evidence for populism to be a driving force of behavioural differences in the context of the pandemic is mixed at best. Since the Brexit referendum and Trump's Presidential election there has been a relentless focus on populism as a key explanatory ingredient for many political and social events. Our study calls into question whether there is indeed systematic evidence for populism to play such a key role in explaining compliance with social distancing measures during the COVID-19 pandemic.

# 8 Appendix

## 8.1 Additional figures

### 8.1.1 Electoral maps

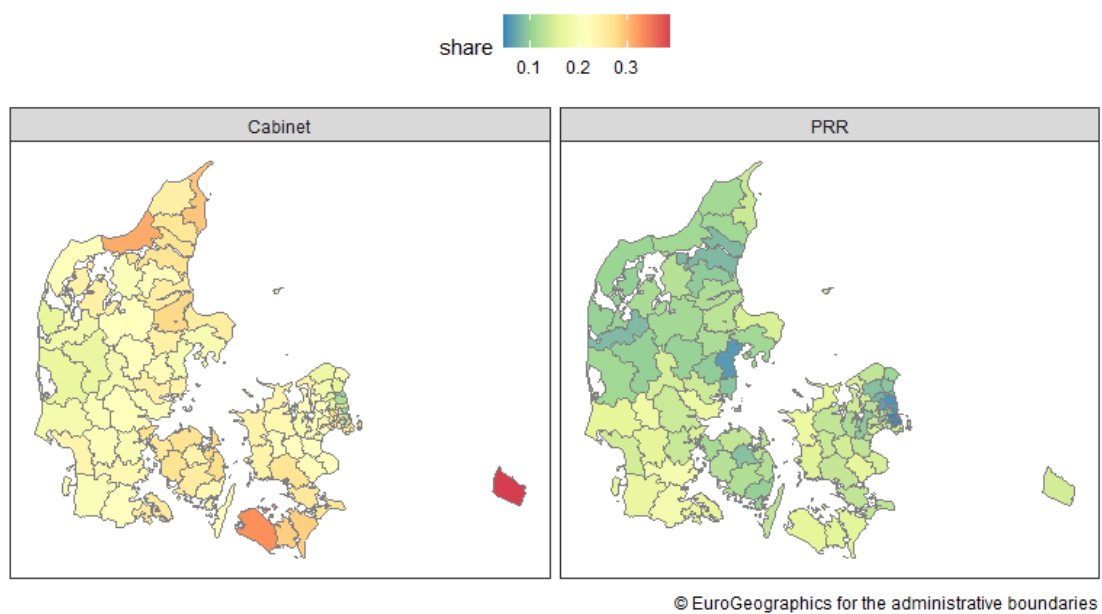

*Figure 8.1: DK: Cabinet (left) and PRR (right) parties' vote shares by NUTS3, standardised*

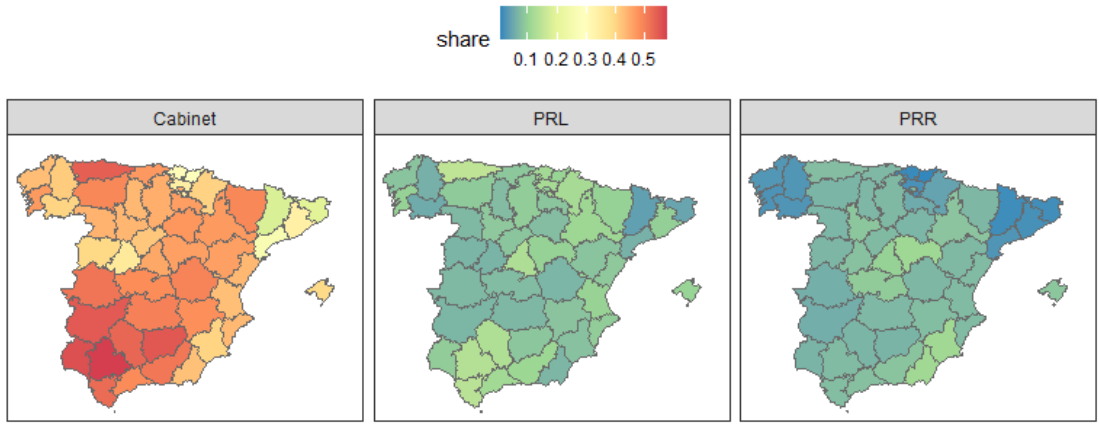

*Figure 8.2: ES: Cabinet (left), PRL (centre), and PRR (right) parties' vote shares by NUTS3, standardised*

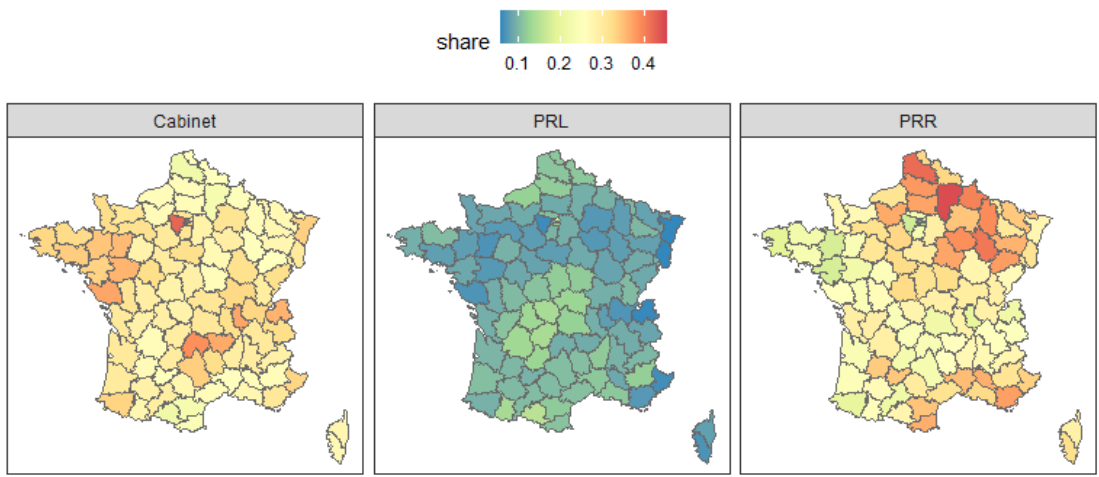

*Figure 8.3: FR: Cabinet (left), PRL (centre), and PRR (right) parties' vote shares by NUTS3, standardised*

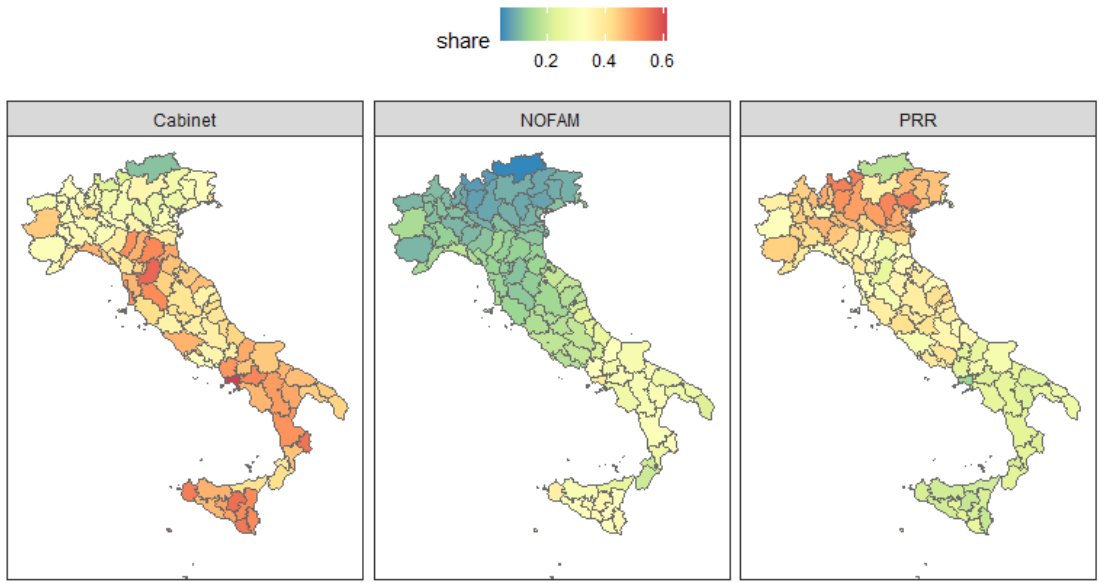

*Figure 8.4: IT: Cabinet (left), NOFAM (centre), and PRR (right) parties' vote shares by NUTS3, standardised*

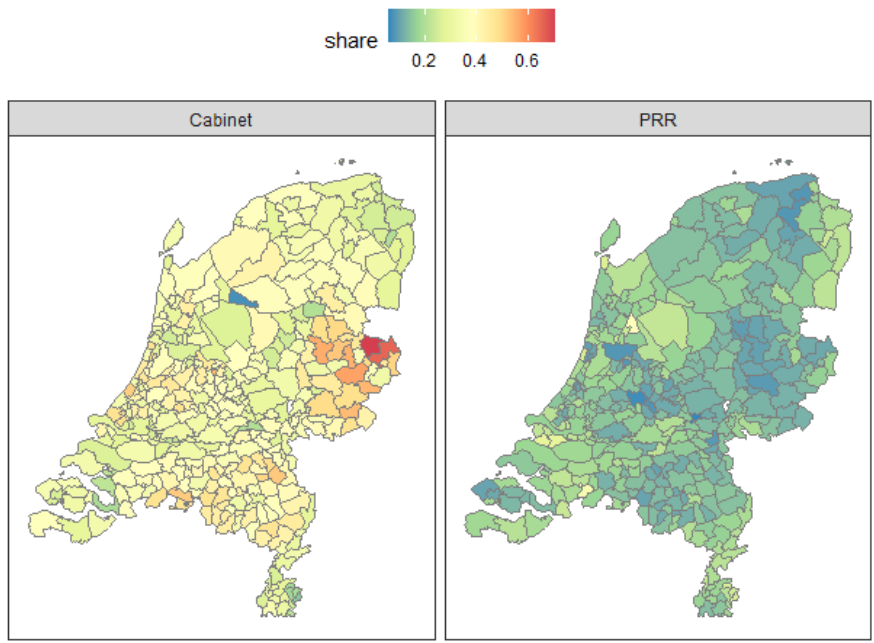

*Figure 8.5: NL: Cabinet (left) and PRR (right) parties' vote shares by NUTS3, standardised*

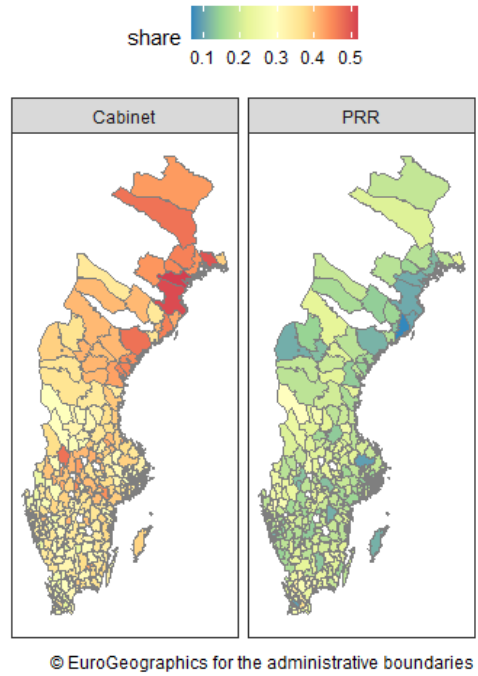

*Figure 8.6: SE: Cabinet (left) and PRR (right) parties' vote shares by NUTS3, standardised*

### *8.1.2* **Mobility trends**

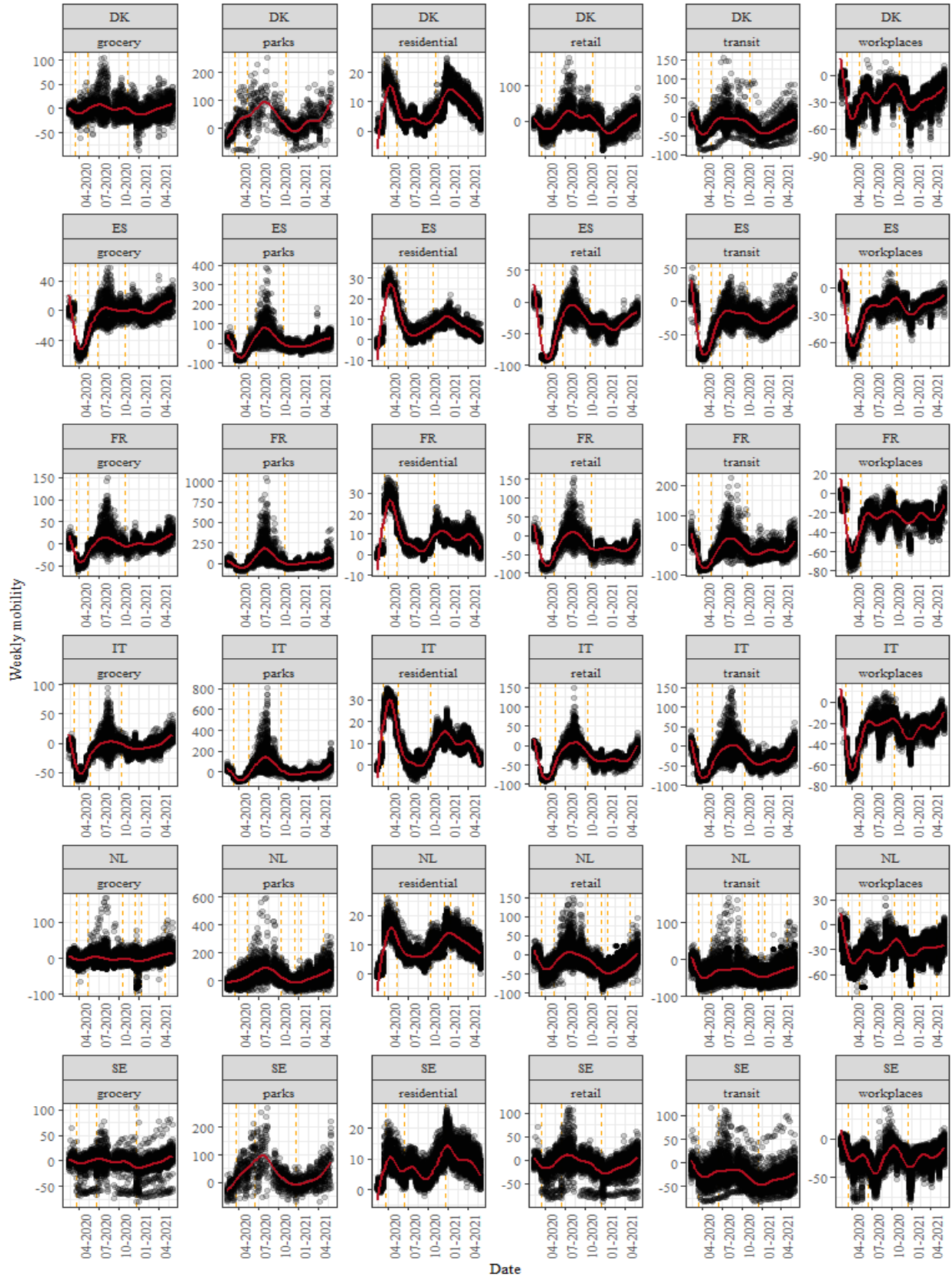
44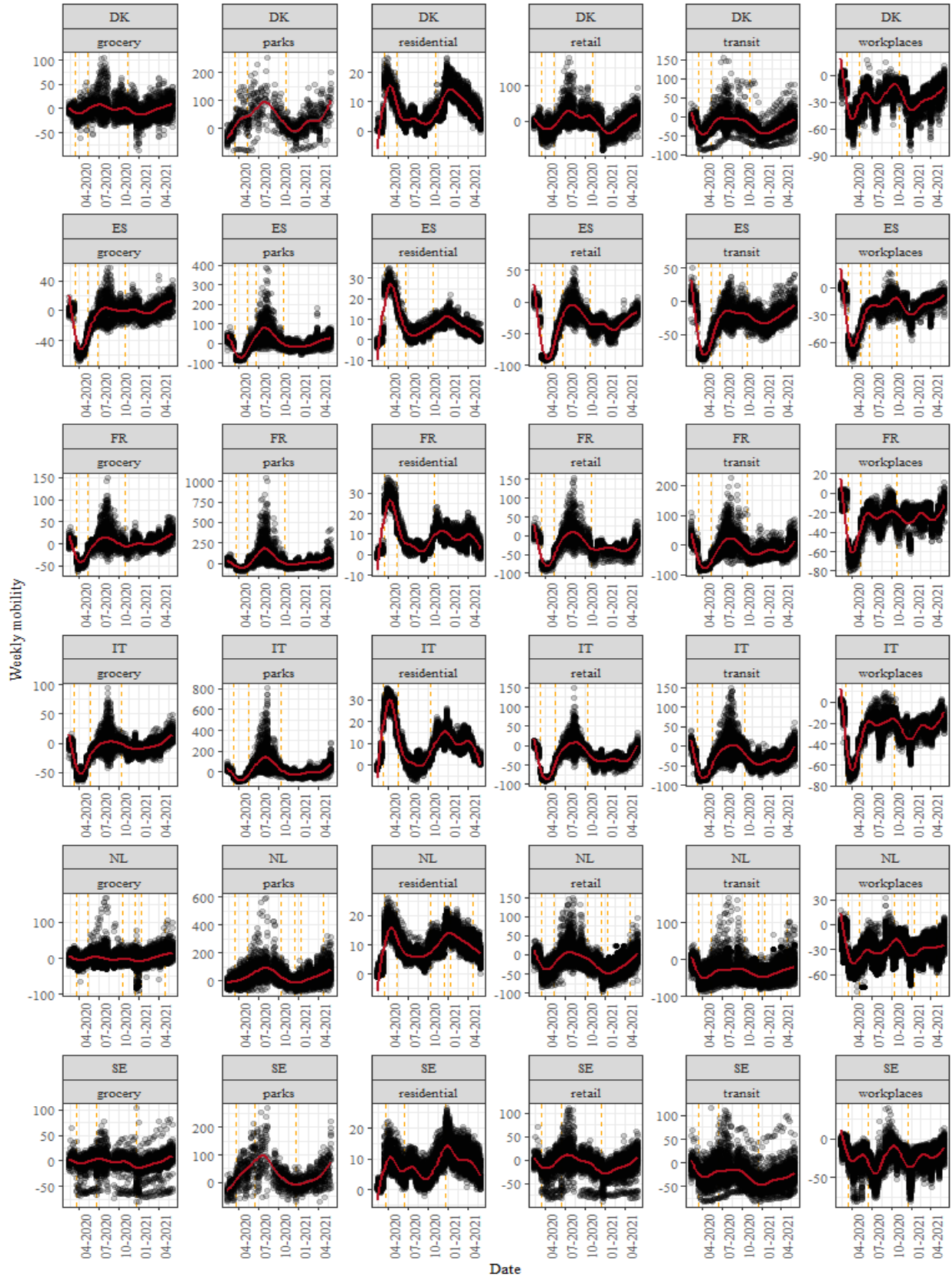

*Figure 8.7: Weekly aggregated mobility, by country and type. Dots are either NUTS3 (ES, FR, IT) or LAUs (DK, NL, SE)*

### 8.1.3 Mobility trends and parties' national shares

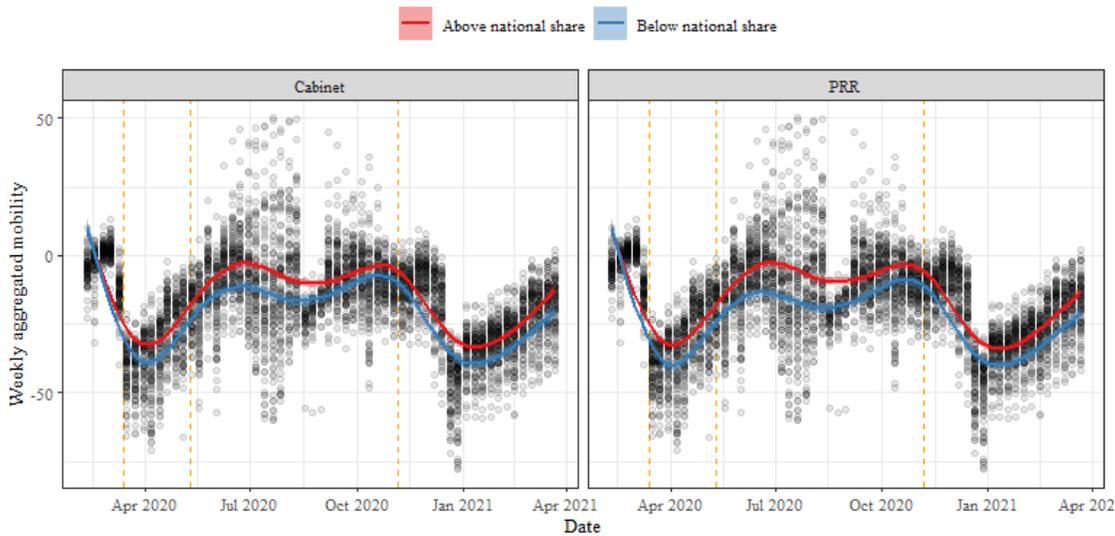

*Figure 8.8: DK: Mobility trends and smoother signalling when regions where above or below the nation-wide governing parties' (left) and PRR (right) share of votes. Dotted line represent noteworthy policy changes*

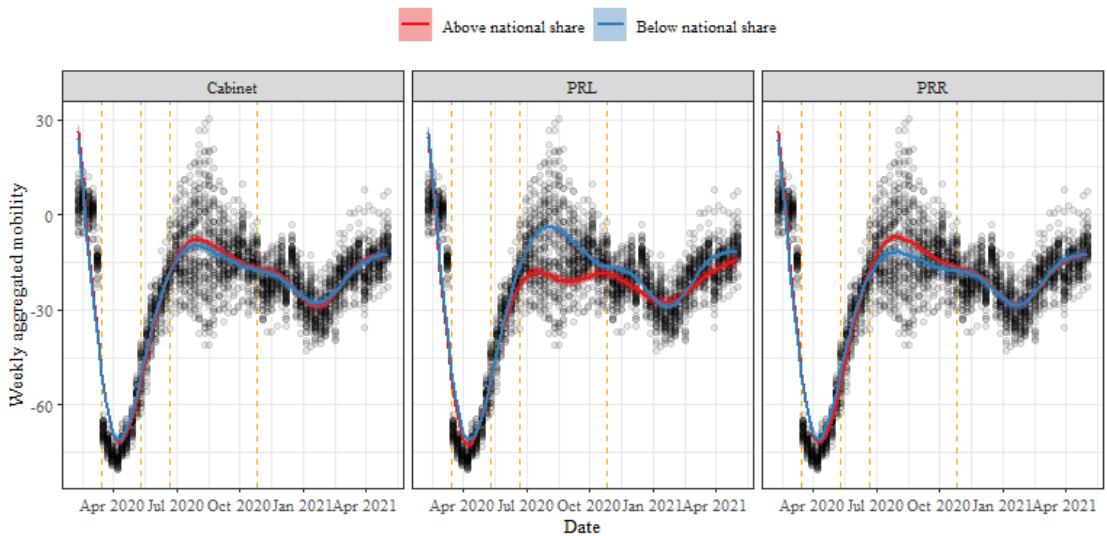

*Figure 8.9: ES: Mobility trends and smoother signalling when regions where above or below the nation-wide governing parties' (left), PRL (centre), and PRR (right) share of votes. Dotted line represent noteworthy policy changes*

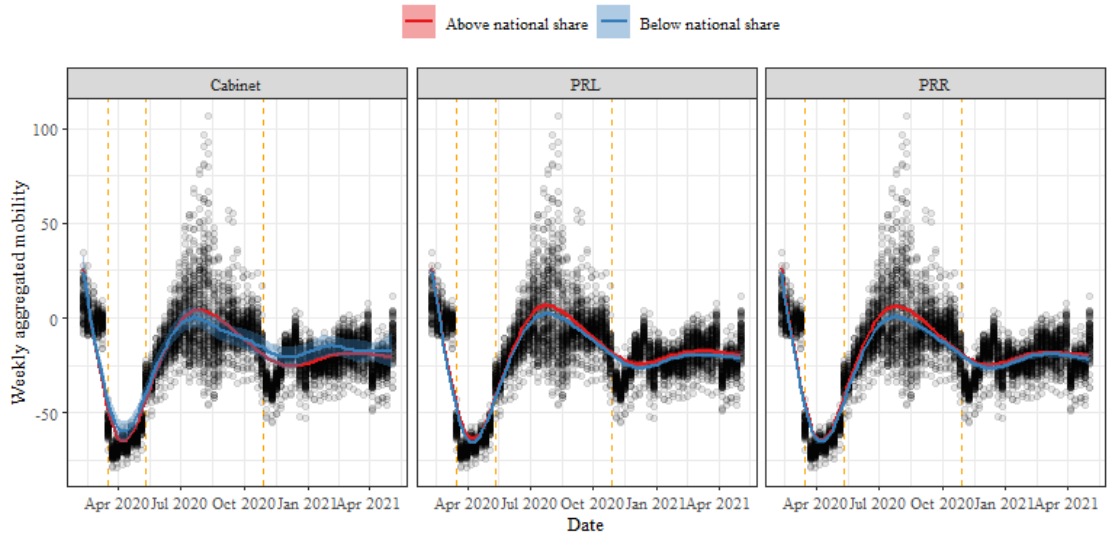

*Figure 8.10: FR: Mobility trends and smoother signalling when regions where above or below the nation-wide governing parties' (left), PRL (centre), and PRR (right) share of votes. Dotted line represent noteworthy policy changes*

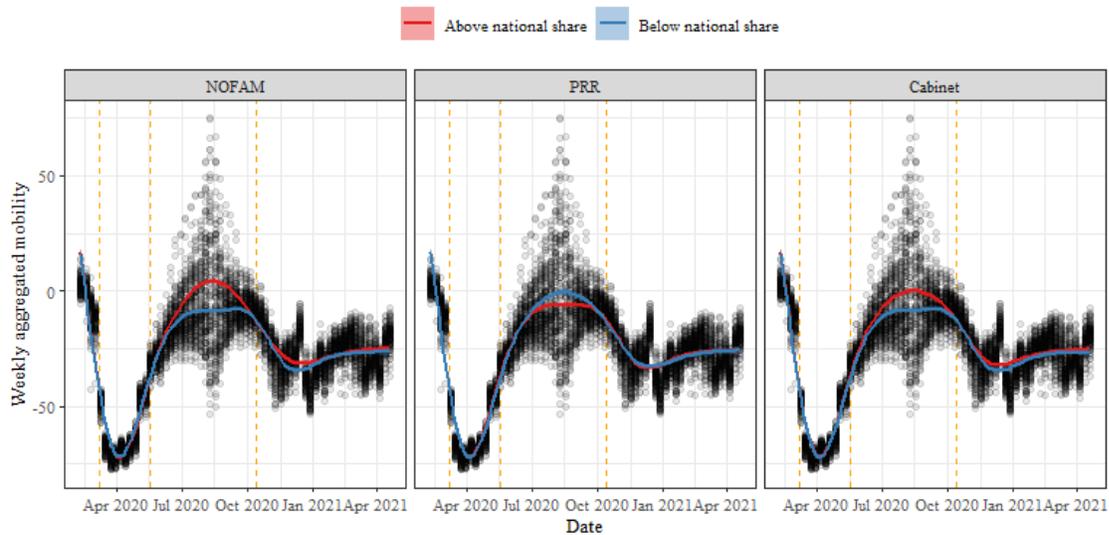

*Figure 8.11: IT: Mobility trends and smoother signalling when regions where above or below the nation-wide governing parties' (left), NOFAM (centre), and PRR (right) share of votes. Dotted line represent noteworthy policy changes*

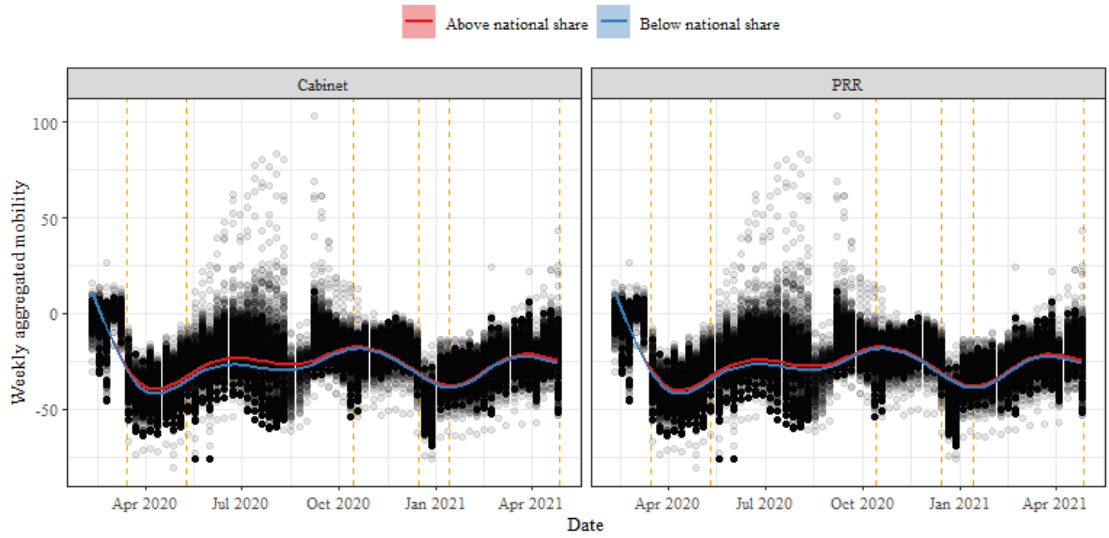

*Figure 8.12: NL: Mobility trends and smoother signalling when regions where above or below the nation-wide governing parties' (left) and PRR (right) share of votes. Dotted line represent noteworthy policy changes*

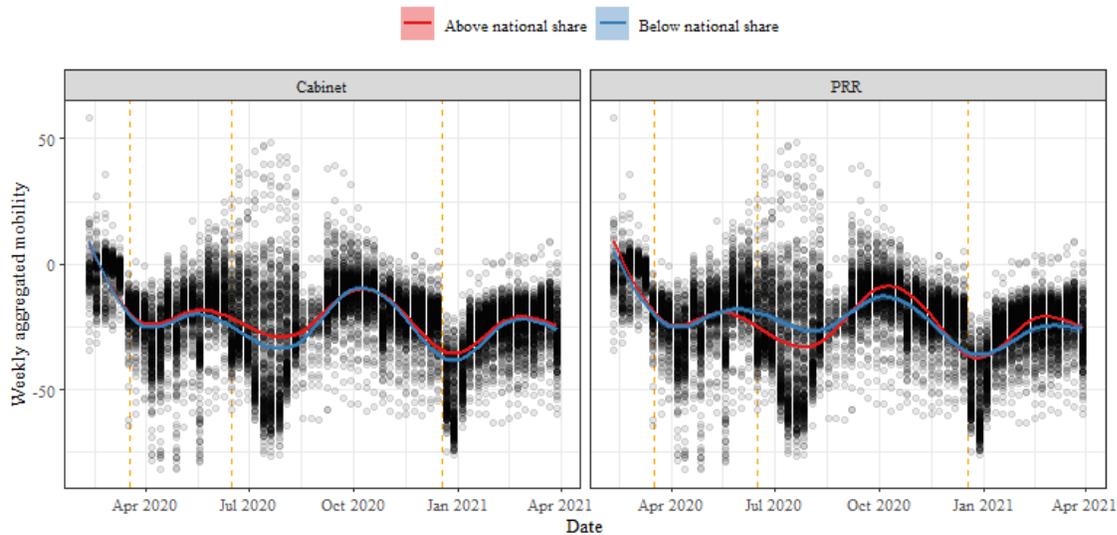

*Figure 8.13: SE: Mobility trends and smoother signalling when regions where above or below the nation-wide governing parties' (left) and PRR (right) share of votes. Dotted line represent noteworthy policy changes*

### 8.1.4 Average mobility indicator: baseline models

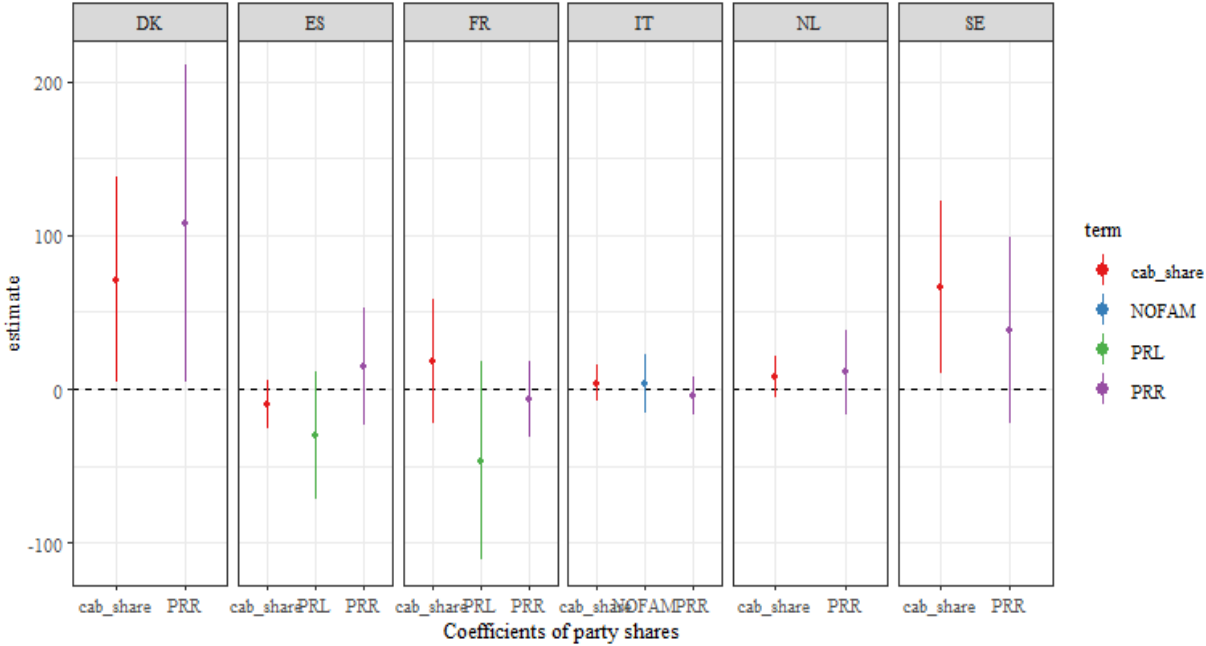

*Figure 8.14: Retail mobility indicator regressed on party share and other covariates, by country*

*8.1.5* **Average mobility indicator: predictions from interactions featuring pandemic periods and parties' vote share**

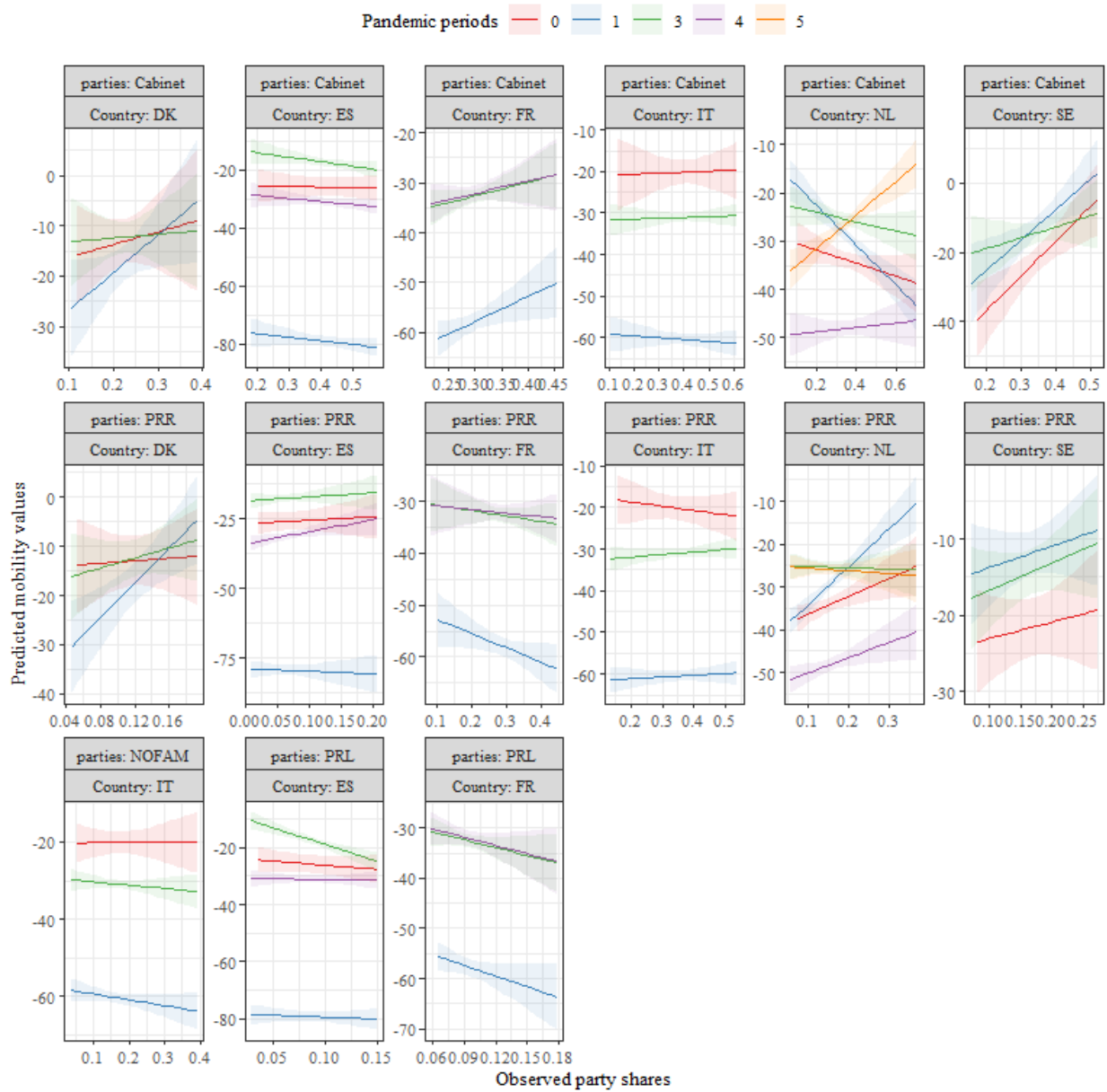

*Figure 8.15: Predicted values of retail mobility, at different levels of parties' share, by different waves and countries.*

*8.1.6* **Weekly coefficients**

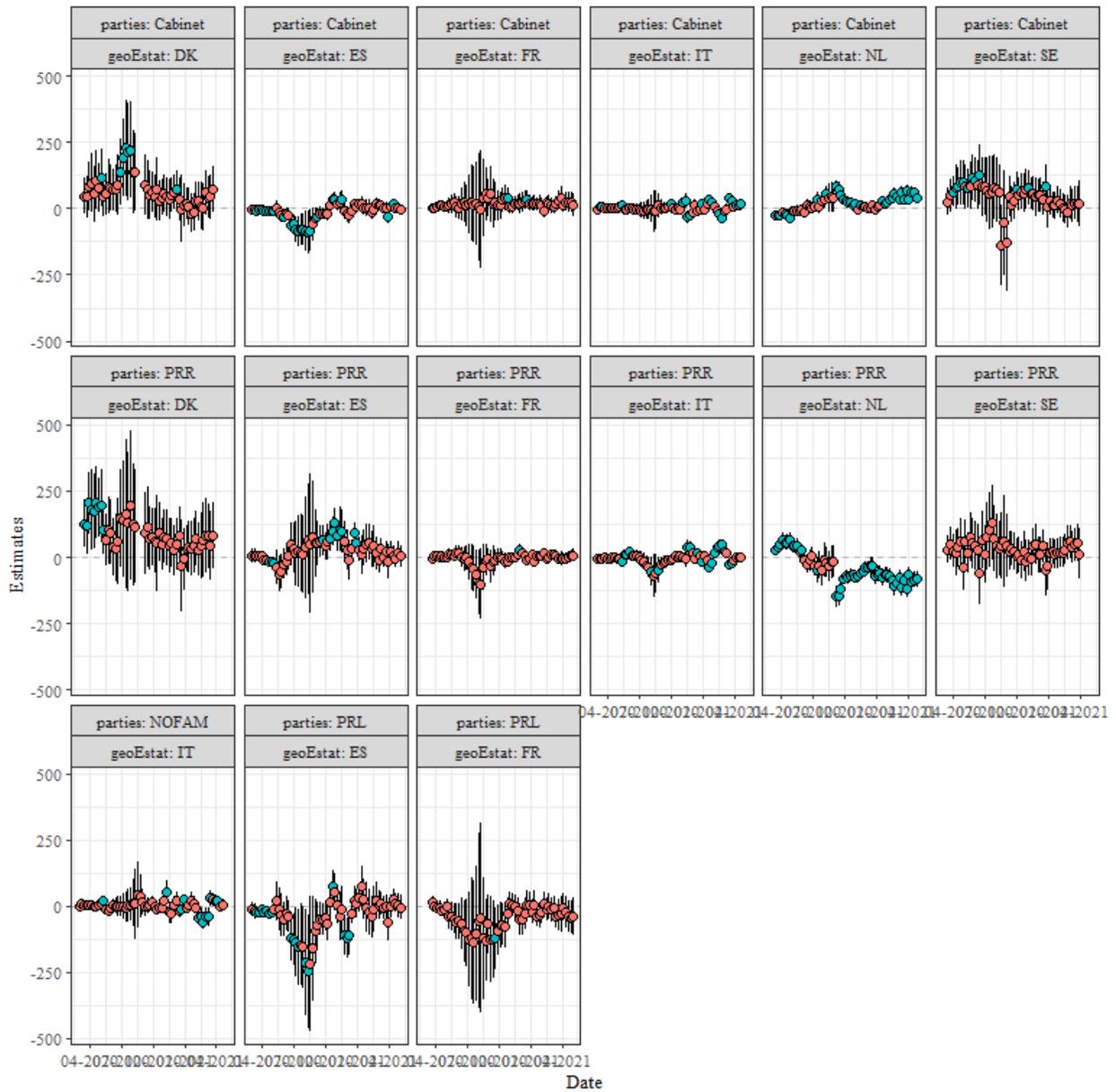

*Figure 8.16: Retail mobility: weekly coefficients by country and party family. Dots are coloured depending on whether p<0.05 (blue) or not (red).*

## *8.2* **Additional tables**

*Table 1. Summary table*

| Characteristic | N = 999,609[1] |
|---|---|
| geoEstat | |
|   DK | 55,410 |
|   ES | 30,895 |
|   FR | 65,346 |
|   IT | 75,446 |
|   NL | 618,618 |
|   SE | 153,894 |
| med_si | 58.12 (18.05) |
|   (Missing) | 70,311 |
| cases_wk_100k | 136.53 (129.74) |
|   (Missing) | 63,417 |
| mean_wkmob_aggr | -15.15 (29.53) |
|   (Missing) | 201,227 |
| mean_wkmob | -9.53 (35.19) |
|   (Missing) | 437,273 |
| nama_gdp | 35,092.00 (10,071.55) |
|   (Missing) | 24,307 |
| old_dep1 | 0.31 (0.06) |
|   (Missing) | 24,307 |
| demodens | 552.04 (815.77) |
|   (Missing) | 24,307 |
| share_empl | 0.43 (0.08) |
|   (Missing) | 178,201 |
| PRR | 0.17 (0.08) |
|   (Missing) | 24,312 |
| PRL | 0.04 (0.03) |
|   (Missing) | 24,312 |
| NOFAM | 0.04 (0.05) |
|   (Missing) | 24,312 |
| cab_share | 0.34 (0.07) |
|   (Missing) | 24,312 |

[1]n; Mean (SD)

**Regressions**

As mentioned above in the Method Section, we employ mainly two different models, namely a multilevel one and a OLS. We report below the equations for both models in the case of Denmark (as the model is the same for all other countries with the exception of Sweden, for which we do not have information regarding employment by economic sectors).

The multilevel model reads

$$
\begin{aligned}
\text{mean\_wkmob\_aggr}_i &\sim N(\mu, \sigma^2) \\
\mu &= \alpha_{j[i],k[i]} + \beta_1(\text{cases\_N3\_wk\_100k\_scale}) + \\
&\quad \beta_2(\text{med\_si\_s}) + \beta_3\big(\text{lag}(\text{cases\_N3\_wk\_100k\_scale})\big) + \\
&\quad \beta_4\big(\text{lag}(\text{med\_si\_s})\big) + \beta_5(\text{wave\_event}_1) + \\
&\quad \beta_6(\text{wave\_event}_2) + \beta_7(\text{wave\_event}_3) \\
\alpha_j &\sim N\left(\gamma_0^\alpha + \gamma_1^\alpha(\text{cab\_share}) + \gamma_2^\alpha(\text{cab\_share} \times \text{wave\_event}_1) + \gamma_3^\alpha(\text{cab\_share} \times \text{wave\_event}_2) + \gamma_4^\alpha(\text{cab\_share} \times \text{wave\_event}_3), \sigma_{\alpha_j}^2\right), \\
&\quad \text{for lau\_code:nuts3 } j = 1, \ldots, J \\
\alpha_k &\sim N\left(\gamma_0^\alpha + \gamma_1^\alpha(\text{demodens\_log\_s}) + \gamma_2^\alpha(\text{gdp\_log\_s}) + \gamma_3^\alpha(\text{old\_dep1\_s}) + \gamma_4^\alpha(\text{share\_empl}), \sigma_{\alpha_k}^2\right), \text{ for nuts3 } k = 1, \ldots, K
\end{aligned}
$$

The OLS models include the same information as the multilevel models, with a few changes. First, we drop the Stringency Index, as that is constant across all geographical units, being a national measurement. The we include only the lag of confirmed cases at NUTS3 level, and not also the confirmed week in the same week in which the mobility is measured, as we do in the multilevel models. Finally, we do not include any interactions with pandemic periods, as the model loop over weeks. reads

$$
\begin{aligned}
\text{mean\_wkmob\_aggr} &= \alpha + \beta_1(\text{gdp\_log\_s}) + \beta_2(\text{old\_dep1\_s}) + \\
&\quad \beta_3(\text{demodens\_log\_s}) + \beta_4(\text{lag\_cases\_N3\_wk\_100k\_scale}) + \beta_5(\text{PRR}) + \\
&\quad \epsilon
\end{aligned}
$$

*8.2.1 Baseline multilevel models*

The main variables of interest - namely political parties' vote shares at either NUTS3 or LAU level - show a very mixed pattern of relationships with mobility, as anticipated before (Figure 8.17).[15] Leaving aside heuristics related to statistical significance, the direction of the coefficients are both positive and negative in the six countries for the same party family. For instance, the PRR party share includes 0 in ES, FR, SE, is positive in NL and DK, and negative in IT. The mixed picture regarding the direction of the relationships between PRR vote share and mobility applies also to the relationship between both PRL and cabinet vote share and mobility. One thing to notice is that PRL coefficients tend to have larger confidence intervals, which is connected to the overall law shares of these parties and thus the fact that they are simply not represented in many geographical entities.

If we turn to retail mobility indicator (Figure 8.14), the picture changes, with most coefficients now crossing 0. This signals the importance of data choices (such as what kind of mobility to consider, or whether or not to aggregate it) on substantive remarks.

Going back to our initial hypotheses 1a and 1b, there seems to be little cross-national evidence that higher shares of votes for cabinet or populist parties correspond to, respectively, greater reductions and increases in mobility.

---

[15] Regression equations and tables are all in the Appendix

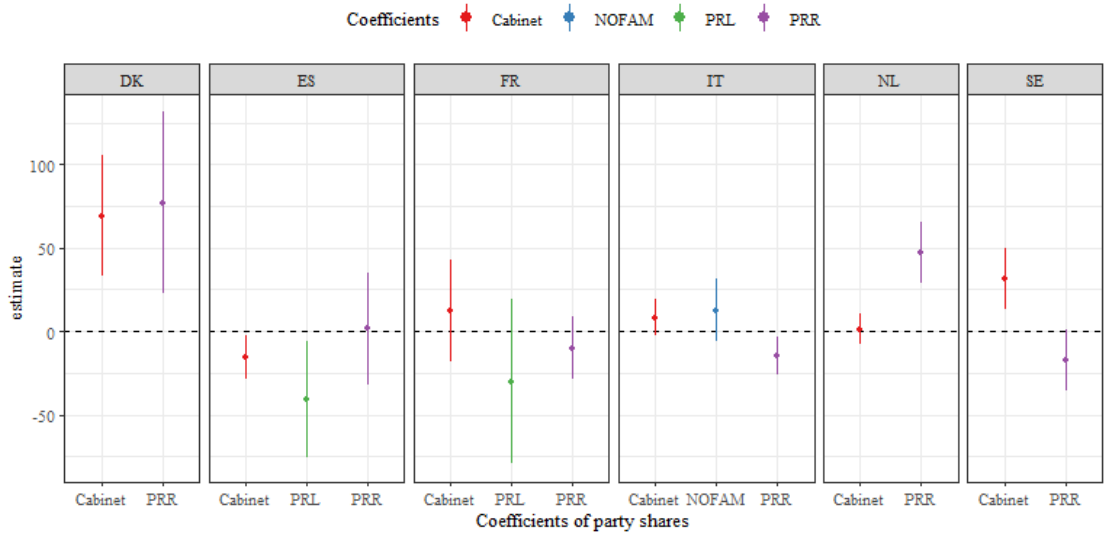

*Figure 8.17: Normalised mobility indicator regressed on party share and other covariates, by country*

While these results are informative, they are also somewhat partial. Indeed, they assume that the relationships between vote share in a region do not vary over time. However, in the context of a protracted crisis, it is possible that the differences captured by variations in party shares across region may become more or less important over time.

### 8.2.2 Regression tables: covariates

*Table 2. Aggregate mobility on covariates*

|  |  | DK | ES | FR | IT | NL | SE |
|---|---|---|---|---|---|---|---|
| (Intercept) |  | 15.39 | -29.45 | -30.21 | -34.69 | -30.76 | -22.43 |
|  |  | [-36.42, 67.20] | [-35.16, -23.75] | [-43.46, -16.97] | [-43.63, -25.76] | [-38.05, -23.47] | [-23.31, -21.54] |
| demodens_log_s |  | -9.34 | -1.13 | 1.53 | -1.04 | -2.27 | 2.38 |
|  |  | [-17.04, -1.64] | [-2.89, 0.62] | [-0.36, 3.42] | [-1.95, -0.13] | [-3.30, -1.23] | [0.85, 3.91] |
| gdp_log_s |  | 3.48 | -0.47 | -2.31 | -3.13 | -0.73 | 0.98 |
|  |  | [-0.76, 7.72] | [-1.55, 0.61] | [-3.90, -0.72] | [-4.09, -2.16] | [-1.80, 0.34] | [-0.64, 2.61] |
| old_dep1_s |  | 2.00 | -0.17 | 4.67 | 2.11 | -0.98 | 2.52 |
|  |  | [-0.53, 4.53] | [-1.38, 1.04] | [3.29, 6.05] | [1.17, 3.05] | [-2.15, 0.19] | [0.28, 4.76] |
| share_empl |  | -68.65 | 21.33 | 27.90 | 21.01 | 17.88 |  |
|  |  | [-177.73, 40.44] | [-0.98, 43.64] | [-1.19, 56.99] | [4.92, 37.09] | [1.03, 34.74] |  |
| cases_N3_wk_100k_scale |  | -1.08 | 0.24 | -0.13 | 0.30 | 0.89 | 0.33 |
|  |  | [-2.44, 0.29] | [-0.61, 1.09] | [-0.62, 0.37] | [-0.17, 0.78] | [0.10, 1.68] | [-1.54, 2.21] |
| med_si_s |  | -13.05 | -12.68 | -21.71 | -14.39 | -5.98 | -5.09 |
|  |  | [-15.17, -10.93] | [-13.64, -11.72] | [-22.74, -20.68] | [-15.45, -13.33] | [-7.22, -4.73] | [-6.54, -3.65] |
| lag(cases_N3_wk_100k_scale) |  | -1.02 | 0.17 | -0.73 | -0.04 | 0.06 | -1.51 |
|  |  | [-2.38, 0.35] | [-0.68, 1.03] | [-1.22, -0.23] | [-0.51, 0.44] | [-0.73, 0.85] | [-3.39, 0.37] |
| lag(med_si_s) |  | 4.25 | 0.19 | -0.92 | -12.15 | -0.91 | -0.35 |
|  |  | [2.15, 6.34] | [-0.76, 1.15] | [-1.93, 0.08] | [-13.20, -11.10] | [-2.15, 0.34] | [-1.80, 1.09] |
| sd__(Intercept) | nuts3 | 1.55 | 2.78 | 4.26 | 3.95 | 1.08 | 0.00 |
|  | lau_code:nuts3 | 7.87 |  |  |  | 6.59 | 7.26 |
| sd__Observation | Residual | 11.76 | 16.33 | 12.32 | 12.52 | 12.60 | 12.36 |
| AIC |  | 40967.9 | 28473.1 | 44723.6 | 48512.5 | 503315.0 | 118510.8 |
| BIC |  | 41046.7 | 28540.4 | 44796.7 | 48586.5 | 503423.7 | 118594.6 |
| Log.Lik. |  | -20471.956 | -14225.543 | -22350.818 | -24245.275 | -251645.499 | -59244.405 |
| REMLcrit |  | 40943.91 | 28451.09 | 44701.64 | 48490.55 | 503291.00 | 118488.81 |

*Table 3. Retail mobility on covariates*

|  |  | DK | ES | FR | IT | NL | SE |
|---|---|---|---|---|---|---|---|
| (Intercept) |  | 53.26 | -38.36 | -35.95 | -42.91 | -24.14 | -9.51 |
|  |  | [-37.69, 144.21] | [-44.93, -31.79] | [-53.33, -18.58] | [-52.04, -33.79] | [-33.99, -14.29] | [-12.30, -6.73] |
| demodens_log_s |  | -12.34 | -0.69 | 1.32 | -0.27 | -0.12 | 9.58 |
|  |  | [-26.25, 1.57] | [-2.71, 1.33] | [-1.15, 3.80] | [-1.19, 0.66] | [-1.52, 1.27] | [4.72, 14.45] |
| gdp_log_s |  | 3.54 | -0.40 | -3.70 | -4.30 | -2.19 | -0.61 |
|  |  | [-4.25, 11.33] | [-1.65, 0.84] | [-5.79, -1.62] | [-5.28, -3.31] | [-3.66, -0.73] | [-5.42, 4.19] |
| old_dep1_s |  | 3.34 | -0.99 | 3.67 | 1.82 | -1.18 | 6.03 |
|  |  | [-0.82, 7.50] | [-2.38, 0.41] | [1.86, 5.49] | [0.86, 2.78] | [-2.74, 0.38] | [-1.00, 13.07] |
| share_empl |  | -115.84 | 15.51 | 23.31 | 29.00 | 3.53 |  |
|  |  | [-307.62, 75.94] | [-10.19, 41.20] | [-14.86, 61.47] | [12.57, 45.42] | [-19.29, 26.35] |  |
| cases_N3_wk_100k_scale |  | -2.89 | 1.74 | -0.91 | 0.18 | -1.94 | 1.07 |
|  |  | [-5.63, -0.15] | [0.70, 2.77] | [-1.51, -0.32] | [-0.42, 0.79] | [-3.08, -0.80] | [-4.41, 6.55] |
| med_si_s |  | -15.43 | -17.29 | -29.49 | -19.95 | -19.01 | -5.64 |
|  |  | [-19.62, -11.24] | [-18.45, -16.13] | [-30.73, -28.26] | [-21.29, -18.61] | [-20.79, -17.22] | [-8.97, -2.31] |
| lag(cases_N3_wk_100k_scale) |  | -3.30 | -3.12 | -2.05 | -1.25 | -2.42 | -7.25 |
|  |  | [-6.04, -0.56] | [-4.15, -2.09] | [-2.65, -1.46] | [-1.85, -0.65] | [-3.56, -1.27] | [-12.73, -1.78] |
| lag(med_si_s) |  | 6.98 | 0.65 | -0.67 | -16.99 | 4.13 | 4.29 |
|  |  | [2.84, 11.11] | [-0.51, 1.81] | [-1.88, 0.54] | [-18.32, -15.66] | [2.36, 5.90] | [0.96, 7.62] |
| sd__(Intercept) | nuts3 | 0.00 | 3.11 | 5.65 | 3.85 | 1.07 | 2.57 |
|  | lau_code:nuts3 | 13.21 |  |  |  | 8.38 | 12.85 |
| sd__Observation | Residual | 21.54 | 19.78 | 14.78 | 15.86 | 15.66 | 14.92 |
| AIC |  | 34283.7 | 29757.7 | 46583.9 | 51275.4 | 205507.4 | 48402.2 |
| BIC |  | 34358.6 | 29825.1 | 46656.9 | 51349.3 | 205604.7 | 48475.6 |
| Log.Lik. |  | -17129.830 | -14867.856 | -23280.959 | -25626.717 | -102741.707 | -24190.101 |
| REMLcrit |  | 34259.66 | 29735.71 | 46561.92 | 51253.43 | 205483.41 | 48380.20 |

### *8.2.3* **Regression tables: political variables**

The following regression tables tabulates the main parameters of interest, by country.

*Table 4. Aggregate mobility on covariates and share of parties in government*

|  |  | DK | ES | FR | IT | NL | SE |
|---|---|---|---|---|---|---|---|
| (Intercept) |  | -23.04 | -23.49 | -31.85 | -39.94 | -31.23 | -33.23 |
|  |  | [-70.72, 24.63] | [-30.99, -15.98] | [-45.71, -17.98] | [-51.27, -28.60] | [-39.25, -23.21] | [-39.46, -27.01] |
| demodens_log_s |  | -5.67 | -1.50 | 1.55 | -1.18 | -2.27 | 3.31 |
|  |  | [-12.44, 1.09] | [-3.21, 0.21] | [-0.34, 3.45] | [-2.11, -0.26] | [-3.31, -1.23] | [1.72, 4.91] |
| gdp_log_s |  | 1.87 | -1.11 | -2.85 | -2.79 | -0.72 | 0.42 |
|  |  | [-1.83, 5.56] | [-2.28, 0.06] | [-4.94, -0.77] | [-3.85, -1.73] | [-1.81, 0.36] | [-1.21, 2.05] |
| old_dep1_s |  | 1.13 | -0.30 | 4.55 | 2.12 | -0.98 | 2.09 |
|  |  | [-1.01, 3.28] | [-1.47, 0.86] | [3.14, 5.97] | [1.19, 3.06] | [-2.15, 0.20] | [-0.12, 4.29] |
| share_empl |  | -20.45 | 23.80 | 23.28 | 24.38 | 17.80 |  |
|  |  | [-115.39, 74.48] | [2.32, 45.29] | [-8.00, 54.57] | [7.76, 41.00] | [0.79, 34.81] |  |
| cases_N3_wk_100k_scale |  | -1.08 | 0.23 | -0.13 | 0.31 | 0.89 | 0.34 |
|  |  | [-2.44, 0.29] | [-0.62, 1.09] | [-0.62, 0.37] | [-0.17, 0.78] | [0.10, 1.68] | [-1.53, 2.22] |
| med_si_s |  | -13.05 | -12.68 | -21.71 | -14.39 | -5.98 | -5.09 |
|  |  | [-15.17, -10.93] | [-13.63, -11.72] | [-22.74, -20.68] | [-15.45, -13.33] | [-7.22, -4.73] | [-6.54, -3.65] |
| lag(cases_N3_wk_100k_scale) |  | -1.02 | 0.17 | -0.73 | -0.03 | 0.06 | -1.52 |
|  |  | [-2.38, 0.35] | [-0.68, 1.02] | [-1.22, -0.23] | [-0.51, 0.44] | [-0.73, 0.85] | [-3.40, 0.36] |
| lag(med_si_s) |  | 4.25 | 0.20 | -0.92 | -12.15 | -0.91 | -0.35 |
|  |  | [2.15, 6.34] | [-0.76, 1.15] | [-1.93, 0.08] | [-13.20, -11.10] | [-2.15, 0.34] | [-1.80, 1.09] |
| cab_share |  | 69.02 | -15.21 | 12.44 | 8.33 | 1.39 | 31.59 |
|  |  | [32.87, 105.16] | [-28.31, -2.11] | [-18.18, 43.07] | [-2.88, 19.53] | [-7.99, 10.76] | [13.59, 49.59] |
| sd__(Intercept) | nuts3 | 0.00 | 2.61 | 4.27 | 3.93 | 1.13 | 0.00 |
|  | lau_code:nuts3 | 7.41 |  |  |  | 6.59 | 7.11 |
| sd__Observation | Residual | 11.76 | 16.33 | 12.32 | 12.52 | 12.60 | 12.36 |
| AIC |  | 40949.3 | 28464.4 | 44717.7 | 48507.1 | 503312.0 | 118494.9 |
| BIC |  | 41034.6 | 28537.9 | 44797.4 | 48587.7 | 503429.7 | 118586.3 |
| Log.Lik. |  | -20461.642 | -14220.222 | -22346.834 | -24241.555 | -251642.976 | -59235.463 |
| REMLcrit |  | 40923.28 | 28440.44 | 44693.67 | 48483.11 | 503285.95 | 118470.93 |

*Table 5. Aggregate mobility on covariates and PRR share*

| | | DK | ES | FR | IT | NL | SE |
|---|---|---|---|---|---|---|---|
| (Intercept) | | 11.18 | -29.58 | -25.91 | -34.97 | -34.85 | -19.06 |
| | | [-36.66, 59.02] | [-36.03, -23.14] | [-41.37, -10.45] | [-43.66, -26.29] | [-42.36, -27.34] | [-22.70, -15.42] |
| demodens_log_s | | -9.32 | -1.12 | 1.40 | -1.00 | -2.76 | 2.70 |
| | | [-16.43, -2.21] | [-2.91, 0.67] | [-0.50, 3.30] | [-1.88, -0.11] | [-3.81, -1.71] | [1.14, 4.25] |
| gdp_log_s | | 3.47 | -0.45 | -2.80 | -2.55 | -0.75 | 0.52 |
| | | [-0.46, 7.39] | [-1.59, 0.68] | [-4.63, -0.97] | [-3.58, -1.52] | [-1.83, 0.34] | [-1.17, 2.22] |
| old_dep1_s | | 1.56 | -0.15 | 4.50 | 2.53 | -1.10 | 2.53 |
| | | [-0.78, 3.89] | [-1.45, 1.16] | [3.08, 5.92] | [1.56, 3.49] | [-2.27, 0.08] | [0.30, 4.76] |
| share_empl | | -79.59 | 21.46 | 24.64 | 30.92 | 10.41 | |
| | | [-180.65, 21.48] | [-1.26, 44.17] | [-5.05, 54.34] | [13.57, 48.28] | [-6.79, 27.61] | |
| cases_N3_wk_100k_scale | | -1.08 | 0.24 | -0.13 | 0.30 | 0.89 | 0.34 |
| | | [-2.45, 0.29] | [-0.61, 1.09] | [-0.62, 0.37] | [-0.17, 0.78] | [0.10, 1.68] | [-1.54, 2.22] |
| med_si_s | | -13.04 | -12.68 | -21.71 | -14.39 | -5.98 | -5.09 |
| | | [-15.16, -10.92] | [-13.64, -11.72] | [-22.74, -20.68] | [-15.45, -13.33] | [-7.23, -4.73] | [-6.53, -3.64] |
| lag(cases_N3_wk_100k_scale) | | -1.01 | 0.17 | -0.73 | -0.03 | 0.06 | -1.52 |
| | | [-2.38, 0.35] | [-0.68, 1.03] | [-1.22, -0.23] | [-0.51, 0.44] | [-0.73, 0.85] | [-3.39, 0.36] |
| lag(med_si_s) | | 4.24 | 0.19 | -0.92 | -12.15 | -0.90 | -0.36 |
| | | [2.14, 6.33] | [-0.76, 1.15] | [-1.93, 0.08] | [-13.20, -11.10] | [-2.15, 0.34] | [-1.80, 1.08] |
| PRR | | 76.86 | 1.55 | -9.91 | -14.98 | 47.34 | -17.34 |
| | | [22.49, 131.22] | [-32.04, 35.14] | [-28.30, 8.48] | [-26.36, -3.60] | [28.88, 65.79] | [-35.53, 0.85] |
| sd__(Intercept) | nuts3 | 1.09 | 2.83 | 4.26 | 3.83 | 1.30 | 0.00 |
| | lau_code:nuts3 | 7.61 | | | | 6.32 | 7.23 |
| sd__Observation | Residual | 11.76 | 16.33 | 12.32 | 12.52 | 12.60 | 12.36 |
| AIC | | 40954.0 | 28467.6 | 44718.2 | 48502.7 | 503286.4 | 118503.0 |
| BIC | | 41039.3 | 28541.0 | 44797.9 | 48583.3 | 503404.1 | 118594.4 |
| Log.Lik. | | -20464.012 | -14221.784 | -22347.103 | -24239.348 | -251630.185 | -59239.518 |
| REMLcrit | | 40928.02 | 28443.57 | 44694.21 | 48478.70 | 503260.37 | 118479.04 |

*Table 6. Aggregate mobility on covariates and PRL or NOFAM share*

| | | DK | ES | FR | IT | NL | SE |
|---|---|---|---|---|---|---|---|
| (Intercept) | | 0.76 | -25.53 | -25.34 | -38.98 | -30.47 | -25.99 |
| | | [-48.62, 50.13] | [-31.92, -19.13] | [-40.79, -9.90] | [-49.95, -28.00] | [-37.93, -23.01] | [-29.03, -22.95] |
| demodens_log_s | | -9.13 | -1.18 | 1.90 | -1.09 | -2.30 | 3.19 |
| | | [-16.06, -2.20] | [-2.86, 0.50] | [-0.08, 3.88] | [-2.00, -0.18] | [-3.35, -1.25] | [1.53, 4.84] |
| gdp_log_s | | 3.11 | -0.28 | -2.68 | -2.34 | -0.72 | 0.53 |
| | | [-0.75, 6.96] | [-1.32, 0.77] | [-4.37, -0.98] | [-3.86, -0.81] | [-1.79, 0.36] | [-1.13, 2.19] |
| old_dep1_s | | 1.60 | -0.33 | 4.96 | 2.36 | -0.94 | 2.48 |
| | | [-0.66, 3.86] | [-1.50, 0.84] | [3.50, 6.42] | [1.35, 3.36] | [-2.13, 0.25] | [0.26, 4.70] |
| share_empl | | -46.39 | 20.18 | 23.57 | 24.75 | 17.97 | |
| | | [-147.17, 54.38] | [-1.20, 41.55] | [-6.31, 53.45] | [7.77, 41.73] | [1.07, 34.87] | |
| cases_N3_wk_100k_scale | | -1.08 | 0.23 | -0.13 | 0.31 | 0.89 | 0.34 |
| | | [-2.44, 0.29] | [-0.62, 1.09] | [-0.62, 0.37] | [-0.17, 0.78] | [0.10, 1.68] | [-1.54, 2.21] |
| med_si_s | | -13.05 | -12.68 | -21.71 | -14.39 | -5.98 | -5.10 |
| | | [-15.17, -10.93] | [-13.63, -11.72] | [-22.74, -20.68] | [-15.45, -13.33] | [-7.22, -4.73] | [-6.54, -3.65] |
| lag(cases_N3_wk_100k_scale) | | -1.02 | 0.17 | -0.73 | -0.03 | 0.06 | -1.51 |
| | | [-2.39, 0.35] | [-0.68, 1.02] | [-1.22, -0.23] | [-0.51, 0.44] | [-0.73, 0.85] | [-3.39, 0.36] |
| lag(med_si_s) | | 4.25 | 0.20 | -0.92 | -12.15 | -0.91 | -0.35 |
| | | [2.15, 6.35] | [-0.76, 1.15] | [-1.93, 0.08] | [-13.20, -11.10] | [-2.15, 0.34] | [-1.79, 1.09] |
| PRL | | 87.73 | -41.09 | -30.09 | | -9.93 | 64.36 |
| | | [-29.33, 204.78] | [-75.89, -6.28] | [-79.52, 19.34] | | [-60.31, 40.44] | [11.78, 116.94] |
| sd__(Intercept) | nuts3 | 0.00 | 2.60 | 4.25 | 3.94 | 1.09 | 0.00 |
| | lau_code:nuts3 | 7.90 | | | | 6.59 | 7.20 |
| sd__Observation | Residual | 11.76 | 16.33 | 12.32 | 12.52 | 12.60 | 12.36 |
| NOFAM | | | | | 12.61 | | |
| | | | | | [-6.30, 31.51] | | |
| AIC | | 40957.9 | 28462.3 | 44715.9 | 48506.5 | 503308.5 | 118498.7 |
| BIC | | 41043.2 | 28535.8 | 44795.6 | 48587.1 | 503426.3 | 118590.0 |
| Log.Lik. | | -20465.964 | -14219.164 | -22345.961 | -24241.237 | -251641.259 | -59237.340 |
| REMLcrit | | 40931.93 | 28438.33 | 44691.92 | 48482.47 | 503282.52 | 118474.68 |

### *8.2.4* Political parties by country

| country | cabinet_party | family | ches_id | party_name_ches | rightwing | radicalright | letfwing | valence |
|---|---|---|---|---|---|---|---|---|
| DK | 1 | 5 | 201 | SD | 0 | 0 | 0 | 0 |
| DK | 0 | 3 | 202 | RV | 0 | 0 | 0 | 0 |
| DK | 0 | 2 | 203 | KF | 0 | 0 | 0 | 0 |
| DK | 0 | 7 | 206 | SF | 0 | 0 | 0 | 0 |
| DK | 0 | 3 | 211 | V | 0 | 0 | 0 | 0 |
| DK | 0 | 6 | 213 | EL | 0 | 0 | 0 | 0 |
| DK | 0 | 1 | 215 | DF | 1 | 1 | 0 | 0 |
| DK | 0 | 3 | 218 | LA | 0 | 0 | 0 | 0 |
| DK | 0 | 7 | 219 | A | 0 | 0 | 0 | 0 |
| ES | 1 | 5 | 501 | PSOE | 0 | 0 | 0 | 0 |
| ES | 1 | 6 | 525 | Podemos | 0 | 0 | 1 | 0 |
| ES | 0 | 2 | 502 | PP | 0 | 0 | 0 | 0 |
| ES | 0 | 3 | 526 | Cs | 0 | 0 | 0 | 0 |
| ES | 0 | 1 | 527 | Vox | 1 | 1 | 0 | 0 |
| FR | 1 | 2 | 609 | RPR; UMP; LR | 0 | 0 | 0 | 0 |
| FR | 1 | 3 | 626 | LREM | 0 | 0 | 0 | 0 |
| FR | 0 | 6 | 601 | PCF | 0 | 0 | 0 | 0 |
| FR | 0 | 5 | 602 | PS | 0 | 0 | 0 | 0 |
| FR | 0 | 7 | 605 | VERTS; EELV | 0 | 0 | 0 | 0 |
| FR | 0 | 1 | 610 | FN; RN | 1 | 1 | 0 | 0 |
| FR | 0 | 6 | 614 | LO-LCR | 0 | 0 | 0 | 0 |
| FR | 0 | 7 | 617 | MEI | 0 | 0 | 0 | 0 |
| FR | 0 | 6 | 627 | FI | 0 | 0 | 1 | 0 |
| FR | 0 | 1 | 628 | DLF | 1 | 0 | 0 | 0 |
| IT | 1 | 5 | 837 | PD | 0 | 0 | 0 | 0 |
| IT | 1 | 5 | 838 | SL; SEL | 0 | 0 | 0 | 0 |
| IT | 1 | 9 | 845 | M5S | 0 | 0 | 0 | 1 |
| IT | 0 | 6 | 803 | RC | 0 | 0 | 0 | 0 |
| IT | 0 | 1 | 811 | LN | 1 | 1 | 0 | 0 |
| IT | 0 | 3 | 813 | RAD | 0 | 0 | 0 | 0 |
| IT | 0 | 2 | 815 | FI | 1 | 0 | 0 | 0 |
| IT | 0 | 8 | 827 | SVP | 0 | 0 | 0 | 0 |
| IT | 0 | 2 | 844 | FDL | 1 | 1 | 0 | 0 |
| NL | 1 | 4 | 1001 | CDA | 0 | 0 | 0 | 0 |
| NL | 1 | 3 | 1003 | VVD | 0 | 0 | 0 | 0 |
| NL | 1 | 3 | 1004 | D66 | 0 | 0 | 0 | 0 |
| NL | 0 | 5 | 1002 | PvdA | 0 | 0 | 0 | 0 |
| NL | 0 | 7 | 1005 | GL | 0 | 0 | 0 | 0 |
| NL | 0 | 6 | 1014 | SP | 0 | 0 | 0 | 0 |
| NL | 0 | 1 | 1017 | PVV | 1 | 1 | 0 | 0 |
| NL | 0 | 7 | 1018 | PvdD | 0 | 0 | 0 | 0 |
| NL | 0 | 9 | 1020 | 50PLUS | 0 | 0 | 0 | 0 |
| NL | 0 | 9 | 1050 | DENK | 0 | 0 | 0 | 0 |
| NL | 0 | 1 | 1051 | FvD | 1 | 1 | 0 | 0 |
| SE | 1 | 5 | 1602 | SAP | 0 | 0 | 0 | 0 |
| SE | 1 | 7 | 1607 | MP | 0 | 0 | 0 | 0 |
| SE | 0 | 6 | 1601 | V | 0 | 0 | 0 | 0 |
| SE | 0 | 11 | 1603 | C | 0 | 0 | 0 | 0 |
| SE | 0 | 3 | 1604 | L | 0 | 0 | 0 | 0 |
| SE | 0 | 2 | 1605 | M | 0 | 0 | 0 | 0 |
| SE | 0 | 4 | 1606 | KD | 0 | 0 | 0 | 0 |
| SE | 0 | 1 | 1610 | SD | 1 | 1 | 0 | 0 |

## *8.3* Periodisation of the pandemic

Identifying some sort of periodisation of the pandemic is instrumental to understand whether the relationship between political preferences and compliance during the pandemic changes over time. Operationally, the ultimate goal for the purposes of this paper is to create a limited number of identifiers to be interacted with our independent variables of interest. That said, there does not seem to be a common definition of what a COVID-19 'wave' has been. When talking about periods, what is generally referred to are increases in cases, deaths, or more restrictive policy measures, or a combination of these. Figures 3.3 and 3.2 have shown how the trendlines relative to confirmed cases and Stringency Index vary across EU member states. To the extent that we define periods based on either pandemic data or the policy response, this enables us to identify stretches of time in which mobility should have not been affected by the pandemic, and that thus we can use as *baselines* to compare subsequent changes in mobility. In this respect, previous studies have generally included dummies for dates marking substantial policy changes Berry et al. (2021) as a strategy to capture changes over time. Considering that the primary interest in such studies is to measure compliance, such choice is reasonable. We replicate this by including in our study noteworthy policy changes in all of the six countries selected here.[16] As any selection of dates inevitably triggers some critiques, we also replicated this exercise by crafting two more periodisations, based local minima of confirmed cases and Stringency Index. Indeed, one could make the case that instead of looking at the severity of policies to establish a periodisation, periods should be defined based on cases to capture the sheer intensity of the pandemic, as policies are just reactions to changing epidemiological circumstances.

If the baseline for normalisation can be defined as *period 0*, *period 1* is that period roughly starting in March 2020 (again, timing varies depending on the country) where the pandemic reached its first climax in several but not all countries as measured by several pandemic-related statistics (confirmed cases, hospitalisations, R0, etc.) as well as the severity of government response. This was followed by a period of relative calm in many European countries (but again not all, as for instance in the case of Spain), roughly coinciding with the summer of 2020. Public behaviour changed sharply during this period, as recorded also in our mobility data (Figure 3.1), and so it is important to properly identify the end of the so-called *first wave* and the start of the *second*, which roughly started during Autumn 2020. The periodisation based on the Oxford Stringency Index takes as baseline the period when the Index was equalling 0. When the Index is no longer 0, we consider the country to have entered *period 1*, and then identify local minima in the trendlines break the observational period into different periods. We follow a similar approach concerning confirmed cases, that is we identify local minima in the number of confirmed cases relative to the national population.

---

[16] See Appendix for a list of these events

To create our periodisation of the pandemic, besides data on the Stringency Index, we get nation-wide confirmed cases from the OxCGRT dataset (Hale et al. 2021). To have a variable which is more comparable across countries, we calculate the new cases for 100,000 population, with Eurostat population data from 2019.

# List of tables